\documentclass{aa}

\usepackage{graphicx}
\usepackage{txfonts}
\usepackage{amsmath}    % Advanced maths commands
\usepackage{amssymb}    % Extra maths symbols
\usepackage[english]{babel}
\usepackage{subfigure}
\usepackage{hyperref}
\usepackage{color,colortbl}
\usepackage{supertabular}
\usepackage{lscape}  % for 'landscape' environment
\usepackage{appendix}
\usepackage[percent]{overpic}
\usepackage{caption}
\usepackage{float}
\usepackage{booktabs}
\usepackage{longtable}
\newcommand{\sg}{S$^4$G}

\renewcommand{\thesubfigure}{(\arabic{subfigure})}

\newcommand{\kpc}{\mathrm{kpc}}

\makeatletter
\renewcommand*\aa@pageof{, page \thepage{} of \pageref*{LastPage}}
\makeatother

\begin{document}

   \titlerunning{Systematic search for tidal features}
   \title{Systematic search for tidal features around nearby galaxies}

   \subtitle{I. Enhanced SDSS imaging of the Local Volume}

   \author{Gustavo Morales
          \inst{1},
          David Mart\'inez-Delgado
          \inst{1},
          Eva K. Grebel
          \inst{1},\\
          Andrew P. Cooper
          \inst{2},
          Behnam Javanmardi
          \inst{3}
          \and
          Arpad Miskolczi
          \inst{4}
          }

   \institute{Astronomisches Rechen-Institut, Zentrum f\"ur Astronomie
              der Universit\"at Heidelberg,\\
              M\"onchhofstr.\ 12--14, 69120 Heidelberg, Germany
              %\email{wuchterl@amok.ast.univie.ac.at}
         \and Institute for Computational Cosmology, Durham University, Science Site, South Road, Durham, DH1 3LE, UK
         \and School of Astronomy, Institute for Research in Fundamental Sciences (IPM), Tehran, 19395-5531, Iran
              %Argelander-Institut für Astronomie, University of Bonn, Auf dem Huegel 71, 53121 Bonn, Germany
         %\email{a.p.cooper@durham.ac.uk}
              %\email{c.ptolemy@hipparch.uheaven.space}
         \and Astronomisches Institut der Ruhr-Universit\"at Bochum, Universit\"atsstr. 150, D-44780 Bochum, Germany
             }

   \date{}

% \abstract{}{}{}{}{}
% 5 {} token are mandatory

  \abstract
  % context heading (optional)
  % {} leave it empty if necessary
   {In hierarchical models of galaxy formation, stellar tidal streams are expected around most, if not all, galaxies. Although these features may provide useful diagnostics of the $\Lambda$CDM model, their observational properties remain poorly constrained because they are challenging to detect and interpret and have been studied in detail for only a sparse sampling of galaxy population. More quantitative, systematic approaches are required. We advocate statistical analysis of the counts and properties of such features in archival wide-field imaging surveys for a direct comparison against results from numerical simulations.}
  % aims heading (mandatory)
   {We aim to study systematically the frequency of occurrence and other observational properties of tidal features around nearby galaxies.  The sample we construct will act as a foundational dataset for statistical comparison with cosmological models of galaxy formation.}
  % methods heading (mandatory)
   {Our approach is based on a visual classification of diffuse features around a volume-limited sample of nearby galaxies, using a post-processing of Sloan Digital Sky Survey (SDSS) imaging optimized for the detection of stellar structure with low surface
brightness.}
  % results heading (mandatory)
   {At a limiting surface brightness of $28\ \mathrm{mag~arcsec^{-2}}$, 14\% of the galaxies in our sample exhibit evidence of diffuse features likely to have arisen from minor merging events. Our technique recovers all previously known streams in our sample and yields a number of new candidates. Consistent with previous studies, coherent arc-like features and shells are the most common type of tidal structures found in this study. We conclude that although some detections are ambiguous and could be corroborated or refuted with deeper imaging, our technique provides a reliable foundation for the statistical analysis of diffuse circumgalactic features in wide-area imaging surveys, and for the identification of targets for follow-up studies.}
  % conclusions heading (optional), leave it empty if necessary
   {}

   \authorrunning{Morales et al.}
   \maketitle
%
%-------------------------------------------------------------------

\section{Introduction}
\label{sec:intro}

In the $\Lambda$ cold dark matter ($\Lambda$CDM) cosmogony, structures grow hierarchically under the influence of gravity through numerous mergers of smaller structures consisting mainly of dark matter (DM) \citep[e.g.,][]{PressSchechter1974,WhiteRees1978,Blumenthaletal1984,Davisetal1985,LaceyCole1993}. Baryonic matter collects in the potential wells of DM halos, then (in sufficiently massive halos) cools and condenses, eventually leading to star formation. State-of-the-art cosmological simulations seek to model the assembly of dark and baryonic mass, star formation, stellar evolution, and so-called `feedback' processes such as supernovae ab initio in order to demonstrate how complex interactions between these processes give rise to the observed diversity of the cosmic galaxy population \citep[e.g.,][]{Kauffmannetal1993,Coleetal2000,Crotonetal2006,Vogelsbergeretal2014,Schayeetal2015}.

In such models, the stellar content of galaxies forms partly in situ through the condensation of gas in the galaxies themselves, and partly through the accretion of stars tidally stripped from other galaxies that they encounter over cosmic time, which may be partially disrupted or have merged completely by the present day \citep[e.g.,][]{searle1978,Abadietal2006, Purcelletal2007}. The assembly histories of galaxies with a stellar mass comparable to the mass of the Milky Way vary widely in these models \citep{GuoWhite2008}. This is supported by observational results from the detailed study of nearby galaxies. Our Milky Way, for instance, appears to have experienced a relatively quiescent merger history \citep{Hammeretal2007}, while its neighbor M31 shows a much more extended stellar structure, including a variety of bright stellar streams with different morphologies \citep[e.g.,][]{Zuckeretal2004,mcconnachie2009,Ibataetal2014,Thomas2017}.

%Models predict that merger rates were higher at higher redshifts \citep[e.g. ][]{Malleretal2006}, and that the merger history depends in characteristic ways on environment \citep[e.g.][]{Gottloeberetal2001}. %Observations also support a variety of mass assembly histories.

A well-established prediction is that most of the mass in stellar halos of present-day Milky Way-mass galaxies was contributed more than $9$ Gyr ago by a few satellites in the mass range of $10^8$ to $10^{9}$~M$_{\odot}$ \citep[e.g.,][]{bullock2005,DeLuciaHelmi2008,cooper10,pillepich2015,Amorisco2017b}. While stars that formed in situ are expected to dominate the stellar mass profiles of galaxies at small galactocentric radii, accreted stars have a much wider range of binding energies and can give rise to stellar halos extending as far as the virial radius of the host DM halo \citep[e.g.,][]{Bullocketal2001,Fontetal2006,Cooper13,Rodriguez-Gomezetal2016}. Long dynamical times in the outer regions of DM halos allow coherent structures formed by tidal stripping, such as streams and shells, to persist for many gigayears \citep{Johnston2001}. Dynamical friction causes the few most massive satellites to deposit their stars at small galactocentric radii, while the outer regions of stellar halos are more likely to consist of material contributed by a number of less massive satellites \citep[e.g.,][]{bullock2005,Amorisco2017}. Owing to their low stellar densities and intrinsically low luminosities, it is hard to determine both the full extent of stellar halos and their contributions to the innermost regions of galaxies. Together with other difficulties, this makes it challenging to constrain stellar fractions observationally ex situ \citep[e.g.,][]{Cooper13,DSouza14,merritt2016,Harmsen17}.

It is more straightforward to detect recent and ongoing accretion events involving satellites that are sufficiently luminous to give rise to bright tidal streams in the outskirts of massive galaxies. A growing number of such features have been detected beyond the Local Group in recent years, and further extragalactic surveys reaching sufficiently low surface brightness have recently been completed or are currently ongoing \citep[e.g.,][]{Schweizer1990,dmd07,dmd08,MouhcineIbata2009,miskolczi11,Ludwigetal2012,Duc15,Okomoto15,merritt2016,Crnojevicetal2016,Spavone17,Harmsen17}.

Mergers between galaxies with very different stellar masses (typically mass ratios of around 1:10 or higher) are often called minor mergers. These generally involve long-period orbits, little orbital decay or angular momentum loss for the less massive galaxy (which we refer to hereafter as the `satellite'), and little disturbance of the central structure of the more massive galaxy. Consequently, thin, coherent stellar tidal streams are a distinctive observable signature of such mergers, more so for less massive, more recently accreted, and kinematically `colder' satellites \citep{johnston08}. Gaseous tidal streams are commonly observed around interacting galaxies and can be easily traced through $21$ cm observations. They usually overlap with the stellar features unless, for instance, ram pressure separates them. Gaseous streams may also be detectable in optical data, for example, through their H$\alpha$ emission or dust content. Pure gaseous streams (such as the trailing Magellanic Stream) are rare, whereas pure stellar streams are more common around massive galaxies following the dispersal of any previously associated gas, or when gas-deficient early-type satellites are disrupted. For a review, see \citet{Duc2013}. In the case of MW-like hosts, we expect minor merger events to be less frequent in the present-day Universe and the coherent structures they generate (such as tidal tails) to persist only for a few billion years before they become undetectable \citep{Wang17}. Observationally, however, the frequency with which such streams occur around MW-like hosts and their distribution of morphologies are poorly constrained.

Over the past decade, the Stellar Tidal Stream Survey (STSS) has carried out an ultra-deep, wide-field imaging exploration of several nearby spiral galaxies, based on data taken with amateur robotic telescopes \citep{dmd08,dmd2009,dmd10,dmd2012,dmd15}. This survey has revealed  striking stellar tidal streams of different morphologies with unprecedented depth and detail. Subsequently, \citet{miskolczi11} developed a search strategy for low-surface brightness tidal structures around a sample of 474 galaxies in the Sloan Digital Sky Survey (SDSS) Data Release 7 archive \citep{Abazajian2009}. The authors calibrated images taken from the SDSS archive and processed them in an automated manner. Searching for possible tidal streams by visual inspection, they found that at least 6\% of their sample showed distinct stream-like features (with a total of 19\% presenting faint features of any kind). This study demonstrated that detecting a meaningful sample of tidal features close to the detection limit of the SDSS images is feasible.

%NOTE: It is necessary to clarify why our study is different than those by Miskolczi11...(Andrew, DMD)  -- APC: yes; it's important to understand if this is was a 'systematic search for streams' or not, if we're going to claim that there is no prior attempt to do what we're doing.

Although considerable progress has been made by \citeauthor{miskolczi11} and other works, studies of structure with low surface brightness in the outskirts of galaxies remain predominantly discovery-driven and qualitative. To enable a meaningful statistical comparison between data with low surface brightness and cosmological models of galaxy formation, two further advances are urgently required: samples with both a well-defined selection function and uniform imaging data, and the development of automated methods to detect and quantify features with low surface brightness.

The majority of existing deep-imaging studies have been targeted at galaxies that are either very nearby or have known features detected in shallower imaging. It is clearly impossible to draw any conclusions about how frequent such features are from these data alone. Furthermore, prior work has focused on structures associated with Milky Way-type galaxies. This definition is subjective;
it typically includes galaxies that are sufficiently bright, morphologically regular, and have late Hubble type. Not only does a subjective selection make it harder to compare one observational sample to another, but it is almost impossible to apply a comparable qualitative selection to models. Currently, even the most sophisticated hydrodynamical simulations do not  reliably reproduce the full range of morphological details that such judgments are based on. Moreover, selection of Milky Way analogues by qualitative criteria will almost certainly result in a wide sampling of the distribution of fundamental quantities such as stellar mass, and a highly incomplete sample at a given stellar mass. It is much more straightforward, and statistically sound, to carry out comparisons in terms of observable quantities that can be robustly constrained in models, stellar mass being the most obvious choice.

Therefore, to make a meaningful comparison between data and models, deep imaging surveys with simple, quantitative selection functions based on fundamental quantities are necessary. Ideally, these would exploit the statistics of brighter circumgalactic features that can be detected in large samples drawn from shallow wide-area surveys, since the expense of targeted deep imaging is often hard to justify for surveys in which substantial numbers of (statistically important) non-detections are to be expected. Low-surface-brightness features are often said to be ubiquitous, but such statements must take into account the brightness of the features and the depth of the observations. A known (and for a given sample, uniform) limit on depth is crucial to make meaningful statements about counts of structures. Finally, since the role of accretion in the galaxy formation process can be investigated through correlations between structure with low surface brightness and other galaxy properties, it will be necessary to examine large numbers of host galaxies of similar mass without restriction to specific morphologies.

In this work, we take a step toward this more systematic approach by making a statistical assessment of the number of features detected in a survey of a volume-, magnitude-, and size-limited sample of nearby Milky Way-mass (as opposed to Milky Way-type) host galaxies. To keep the study consistent, we select our sample on the basis of mass and recessional velocity. We apply a custom image reduction process uniformly to images of each galaxy in our sample from Data Release 10 of the SDSS \citep{ahn14}, reaching a detection limit in surface brightness of approximately $28\ \mathrm{mag~arcsec^{-2}}$.

This paper presents our observational results. In subsequent work, they will form the basis of further investigations into the properties and recent evolution of stellar halos and comparisons with theoretical predictions from $\Lambda$CDM models. Stellar halos are believed to have formed through a series of accretion events occurring over the lifetime of their host galaxies. Debris associated with the most ancient mergers and those with intrinsically faint progenitors is likely to have extremely low surface brightness at the present day (below $30\ \mathrm{mag~arcsec^{-2}}$). The technique we describe here is therefore well suited to studying evidence for more recent ($t_\mathrm{lookback} \sim 4-5\ \mathrm{Gyr}$) interactions and mergers with more massive satellite galaxies, rather than ancient, well-mixed halo components or the contribution of fainter satellites.

%formed much earlier than the epoch being investigated here, and through the coalescence of much less massive satellites than can be explored using these techniques, this means that studies like these can effectively tackle the key question of the presence (or absence) of ongoing or recent minor mergers, but cannot say very much about the detailed accretion history of such stellar halos. Moreover, the extraordinarily low surface brightness levels (well below $30\ \mathrm{mag~arcsec^{-2}}$) of the merging events that are believed to have formed the bulk of the stellar mass in halos mean that we can only detect the more rare, comparatively luminous structures with comparatively high surface brightness, but not the fainter, more common structures.

Several previous surveys of tidal features have been published, albeit with some key differences in sample selection. \citet[][]{Kaviraj2010} focused on early-type galaxies (ETGs) and found that $\sim 18\%$ of their sample exhibited signs of disturbed morphologies (e.g., shells). This sample was also based on SDSS multiband photometry, but combined with the significantly ($\sim2 \mathrm{mag}$) deeper monochromatic images from the SDSS Stripe 82. \citet{Atkinson2013} studied faint tidal features in galaxies with a wide range of morphologies using the wide-field component of the Canada–France–Hawaii Telescope Legacy Survey. Their sample consisted of 1781 luminous galaxies in the magnitude range $15.5 < r < 17.0$. A classification of tidal features according to their morphology (e.g., streams, shells, and tails) was performed, with no major interpretation in terms of their physical origin, especially when distinguishing between major and minor mergers. They found that about 12\% of the galaxies in their sample showed clear tidal features at their highest confidence level. This fraction increased to about 18\% when they included systems with weaker tidal features. The colors and stellar masses of  central galaxies were found to influence these numbers significantly: linear features, shells, and fans were more likely in galaxies with stellar masses $>10^{10.5}\ \mathrm{M_{\odot}}$, and red galaxies were found to be twice more likely to show tidal features than blue galaxies. Table 1 from \citet{Atkinson2013} summarizes an overview of faint substructures studies from earlier work in the literature. We note that no publication attempted a less restricted but still controlled sample, especially focused on a future comparison with state-of-the-art simulations.

Throughout the text, we use the term `overdensity' to refer to any kind of diffuse feature in the processed image that is not obviously the outward continuation of the brighter isophotes of the host galaxy, without making claims regarding their origin or nature (including whether they are real stellar features or are physically associated with the host galaxy). Minor merger signatures, and more specifically, stellar tidal streams, are understood as a particular class of overdensities, arising from stars distributed around the orbit of a current or former satellite, or else a tidal distortion of the host galaxy. In cases where a host galaxy interacts with a companion of comparable mass (typically referred to as a major merger), both may be severely distorted. Our sample contains very few of these non-equilibrium systems, which we exclude from further consideration.

 In Section \ref{sec:data} we describe our sample selection and image post-processing technique. Section \ref{sec:res} presents our results, including the discovery of several new streams and a list of tidal feature candidates for follow-up observations. Section \ref{sec:con} discusses these results and directions for future work. The tables referenced throughout the paper are presented in Appendix \ref{app}.%Throughout the text, the authors use the concept of overdensities to refer to any kind of features in the image that are bright enough to get enhanced by the processing method, irrespective of their nature and origin. Minor merger signatures, and more specifically, stellar tidal streams, are defined as a particular example of overdensities displaying baryonic content in the tidal field, assumed to be physically related to the host galaxy, and tracing stellar material that has been disrupted or stretched along the orbits of the host galaxy or its satellites, but leaving any sort of major interactions out of the picture\footnote{This includes satellites and spheroids, but dismisses galactic collisions and interactions.}. The latter are more commonly related to post-merger events, that will be understood as any gravitational interaction posterior to the formation of the galaxy that may or may not disrupt the target galaxy in the same way minor merging events do. In this way, even if the outcome of post-mergers may emulate detectable tidal features (like warped disks, polar rings or stellar disruptions), they are not tracing galaxy formation in the same vein minor merging do, and thus are ignored when reporting the frequency of stellar tidal streams.

\section{Data}
\label{sec:data}

%Observationally, stellar tidal streams are challenging to detect due their low-surface-brightness. Its stellar composition has a direct impact on their detectability, but in general, streams are believed to be ubiquitous. In other words, while from simulations they are expected to be physically present, observationally we are not detecting all of them, if present at all.
The aim of this work is to compile a catalog of diffuse overdensities to a known, uniform limiting depth around an approximately volume- and mass-limited sample of host galaxies. This will allow us to constrain the rate of occurrence of tidal debris at the present day and hence (in future work) to test predictions for the frequency and effects of low-redshift minor mergers in galaxy formation models. This section describes how we used the Spitzer Survey of Stellar Structure in Galaxies \citep[\sg ,][]{sheth10,Querejeta2015} to select such a sample of host galaxies, and how we processed the SDSS imaging data for these galaxies in order to search for diffuse overdensities.%Note that all of the images used in this article have half a degree by side with North pointing upwards, unless stated otherwise.

%Since satellites in merging processes within the hierarchical paradigm consist mainly of stars, gas and dark matter, they emit most of their thermal radiation in the infrared and optical parts of the electromagnetic spectrum. Thus, infrared studies are ideal as a starting point in terms of detection capabilities, which then can be followed-up by optical surveys.

\subsection{Sample}
The \sg\ is a volume-, magnitude-, and size-limited ($d< 40$ Mpc, $|b| > 30^\circ$, $m_{\mathrm{B_{corr}}} < 15.5$, and $D_{25} > 1'$) survey of 2352 galaxies using the Infrared Array Camera \citep[IRAC, ][]{Fazio2004} of the Spitzer Space Telescope \citep{Werner2004} at 3.6 and 4.5 $\mu m$. The azimuthally averaged surface brightness profiles obtained by \sg\ are typically robust to isophotes at $\mu_{3.6\ \mu m}(AB) \sim 27\ \mathrm{mag~arcsec^{-2}}$, equivalent to a stellar mass surface density of about 1 $\mathrm{M_{\odot}~pc^{-2}}$ \citep{MunozMateos2015}. \sg\ thus provides an appropriate data set for the study of the distribution of stellar mass and structure in the local Universe, and it is complete for galaxies within the volume relevant to our work and for masses greater than $10^{9.2}\ \mathrm{M_{\odot}}$, allowing us to select a statistically representative sample of galaxies whose stellar masses have been measured in a uniform manner.

Our work focuses on the frequency of tidal features around galaxies at and above the stellar mass of the Milky Way, because contemporary cosmological volume simulations can readily resolve these galaxies and their brighter satellites, which give rise to the most conspicuous features. We therefore selected elliptical, spiral, and S0 galaxies (according to the morphological type code $T$ given by \sg) with a lower stellar mass limit of $10^{10} \mathrm{M_{\odot}}$ in the \sg\ catalog. Constraining the sample in stellar mass limits bias when comparing with simulations, because samples can be selected using equivalent criteria in both. We excluded any galaxies in the region of the Virgo cluster
from our parent sample, as defined by the Next Generation Virgo Cluster Survey (NGVS) footprint \citep{munoz14}, in both projected position and line-of-sight distance ($15 < d_\mathrm{L.O.S.}\ (\mathrm{Mpc}) < 18$). The study of diffuse circumgalactic structure in dense environments such as Virgo is complicated by additional tidal forces of the cluster potential acting on the host and its satellites. Virgo is a `rare' system in the context of the volume we study here, and excluding it allowed us to better represent the statistics of lower-mass groups and isolated galaxies. Clusters of mass comparable to Virgo ($\sim10^{14}\ \mathrm{M_{\odot}}$) can easily be identified in simulations, so this does not compromise a straightforward model-data comparison. Moreover, $17$ known major mergers were removed from the final sample\footnote{Targets removed as major mergers are NGC 2798, NGC 3166, NGC 3169, NGC 3190, NGC 3227, NGC 3998, NGC 5953, NGC 5954, NGC 4550, NGC 5774, NGC 5775, NGC 3395, NGC 5194, NGC 5195, NGC 4302, NGC 5566, and NGC 5574.}. We note that \sg\ already excludes targets at low Galactic latitudes ($|b|<30^\circ$), which is appropriate for our purposes because the detection of features with low surface brightness is severely limited by the presence of extended Galactic cirrus, high extinction, and stellar crowding. Finally, since we used SDSS imaging, we also excluded \sg\ galaxies outside the SDSS footprint.

\begin{figure*}
\renewcommand*\thesubfigure{(A\arabic{subfigure})}
  \begin{center}
    \includegraphics[width=\linewidth]{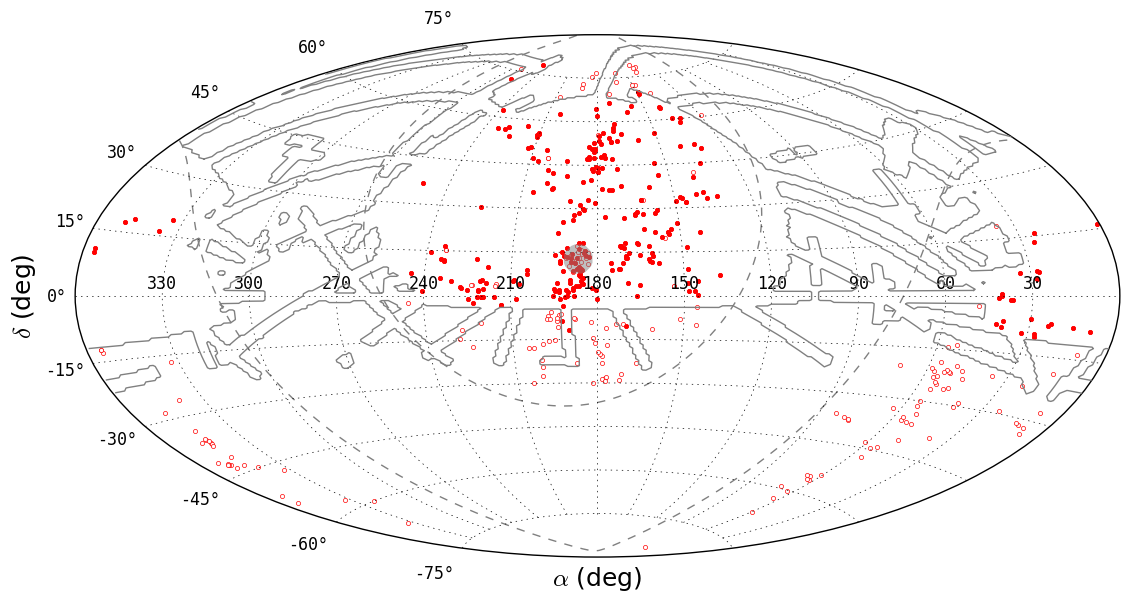}
  \end{center}
  \caption{Aitoff projection of the full \sg\ catalog (empty and filled red circles). The filled red circles mark our selected parent galaxy sample. The dashed lines enclose the Galactic plane area, the solid gray line encompasses the SDSS DR10 footprint, and the solid gray region encompasses the Virgo Cluster area as defined by the NGVS.}
  \label{fig:map}
\end{figure*}

We processed the SDSS images of the targets selected from \sg\ in two individually volume-complete chunks to obtain samples that were feasible for the observing time constraints on follow-up observations of the robotic telescope network used in the STSS, which prioritizes more nearby galaxies. In this first paper we present the results for a Local Volume sub-sample (galaxies selected with a recession velocity lower than 2000 km/s), comprising a total of 297 galaxies. Figure \ref{fig:map} shows the distribution of the parent sample across the sky in celestial coordinates. As shown later, the distance distribution peaks around $20\ \mathrm{Mpc}$, roughly at the boundary of the Local Volume \citep{karachentsev06}. The stellar mass distribution of the sample is limited to the mass range  $10^{10-11} \mathrm{M_{\odot}}$, which follows directly from our selection function. Both mass and distance measurements were taken directly from the \sg\ catalog. Later figures (Section \ref{sec:res}) show the sample distribution morphologically and in inclination angle as well.

\subsection{SDSS data handling and imaging processing}
\label{ss:data_acq}

We used SDSS imaging of the target galaxies selected from \sg\ in order to search for faint features in their surroundings. The SDSS imaging camera worked in drift-scan mode, opening its shutter for extended periods and imaging a continuous strip of the sky \citep{Gunn1998}. This means that, while not very deep, the SDSS imaging survey was able to deliver data with consistently low systematic variations from field to field and excellent flat-fielding. These conditions are critical when searching for extended, diffuse-light features close to the detection limit. The SDSS imaging data also lie in the optical range and have better angular resolution than those of \sg, which is why the latter was used to select the sample, but not for the discovery of stellar substructures. See \citet{Laine2014} for a comprehensive survey of faint structures in the \sg\ images.

For each of the 297 target galaxies selected from \sg, we downloaded and reprocessed the available SDSS DR10 imaging archive data. We followed the procedure described by \citet{miskolczi11}. This consists of four steps: (1) mosaicing of the SDSS images in each bandpass; (2) stacking images in multiple bands to improve the signal-to-noise ratio (S/N), with no weighting applied; (3) two-stage source extraction, including removal of point sources; and (4) Gaussian filtering to enhance features on the scale of interest.

%\begin{figure}
%\renewcommand*\thesubfigure{(A\arabic{subfigure})}
%  \begin{center}
%    \includegraphics[width=\columnwidth]{dm_hist4.png}
%  \end{center}
%  \caption{Distance and mass histograms of the sample, including targets with merging events found around the host galaxy.}
%  \label{fig:hist_dm}
%\end{figure}

%\begin{figure}
%\renewcommand*\thesubfigure{(A\arabic{subfigure})}
%  \begin{center}
%    \includegraphics[width=\columnwidth]{ti_hist2.png}
%  \end{center}
%  \caption{Distribution of the morphology and inclination angle of the sample, including targets with merging events found around the host galaxy.}
%  \label{fig:hist_t}
%\end{figure}

Square mosaics of $30$ arcmin ($4595$ pixels) per side were created in three filters, $g$, $r,$ and $i$, using the automatic script in the SDSS Science Archive Server that can be found online\footnote{\url{dr10.sdss.org/mosaics}} (at $20\ \mathrm{Mpc}$, $30$ arcmin corresponds to approximately $176\ \mathrm{kpc}$). We used these three filters because they have the highest reported sensitivity, and because their combined transmission curve closely resembles the luminance filters used by other observational works, such as \citet{dmd10}. The mosaics of each filter were then stacked using the IRAF task \verb+imcombine+ with the default parameters, in order to improve the S/N of the image. We call this stacked image $I_\mathrm{gri}$.

Our analysis relies heavily on visual inspection of extended, diffuse features in moderately crowded stellar fields. We therefore processed the stacked images with SExtractor \citep{sextractor96} using a two-stage procedure (known as ``hot and cold run", \citealt{caldwell08}) in order to remove the majority of unsaturated point sources while preserving regions of diffuse emission. In Step 1, we extracted all sources covering an area of at least 5 pixels at a signficance of $1.5\sigma$ and saved a FITS file, $I_5$, containing these detections (including the central galaxy, which covers a significant fraction of the image). In the second run, the minimum source area was set to 800 pixels ($\approx 30\ \mathrm{kpc^{2}}$ at $20\ \mathrm{Mpc}$), so that only the central galaxy and any other large objects were detected. We call the corresponding FITS file $I_{800}$. A final image was then created by subtracting $I_{800}$ from $I_5$, that is, the large-scale source(s) from the total detections. The image resulting from this operation contains only compact sources (mainly stars). We call this image file $I_s= I_5 - I_{800}$. Using the IRAF task \verb+imarith+, the sources previously extracted were subtracted from the original stacked image, that is, $I_* = I_\mathrm{gri} - I_s$. Thus, $I_*$ is a stacked image with most of the stars in the field masked out, replaced by the average flux of the neighboring pixels.

\begin{figure*}
\centering
\renewcommand*{\arraystretch}{0}
\begin{tabular}{*{3}{@{}c}@{}}

\includegraphics[draft=false,width=0.33\linewidth]{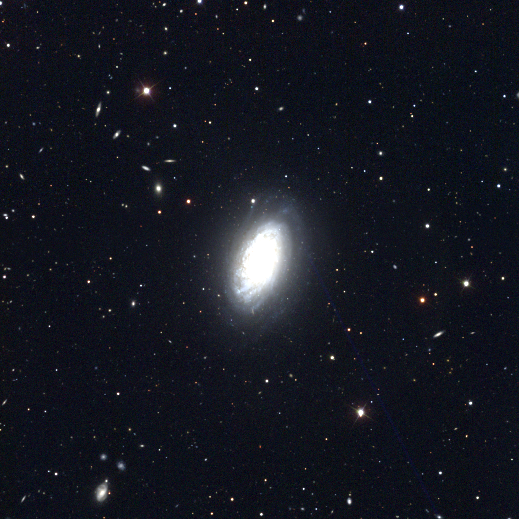}&
\includegraphics[draft=false,width=0.33\linewidth]{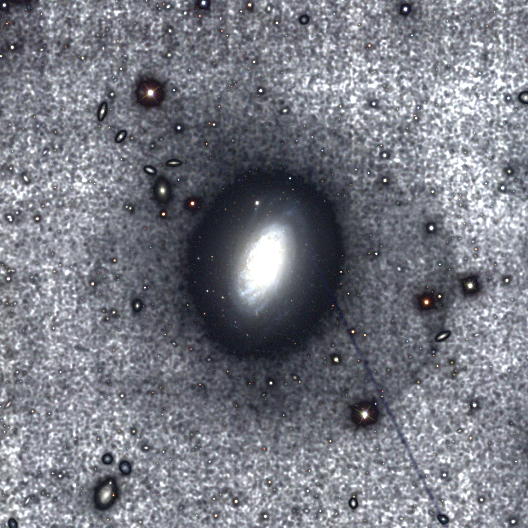}&
\includegraphics[draft=false,width=0.33\linewidth]{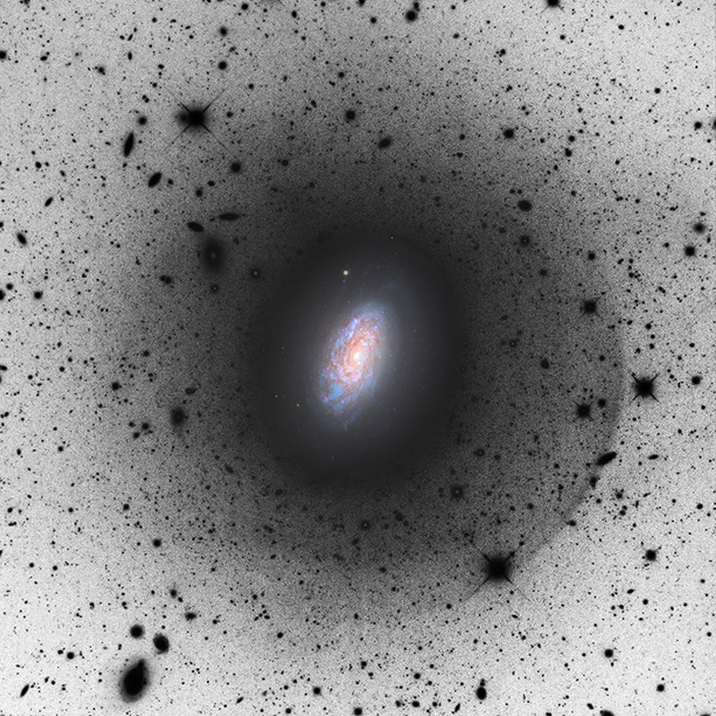}\\

\end{tabular}
\caption{Example of the enhancement technique described in Sec.2.2 for NGC 4414. The field of view for each panel is $\sim$ 30 $\times$ 30 arcminutes (north is up, east to the left). {\it Left panel}: Original SDSS color-composite image from the public archive. {\it Middle panel}: Final result of our image processing with an inverted, stretched grayscale showing the extent of a conspicuous shell of debris southwest of the central galaxy. For illustrative purposes, we have added a color inset of the disk of the galaxy and any stars or other objects that are masked in our analysis. {\it Right panel}: Same field taken from the STSS (Martinez-Delgado et al. in preparation), showing the same overall morphology of this substructure. Credit right panel image: Adam Block.}
\label{fig:proc}
\end{figure*}

To enhance the visibility of faint, extended features, we then applied a circularly symmetric Gaussian filter to $I_*$.  \citet{miskolczi11}
reported that other possible filter types are available in IRAF, including \verb+adaptive+ and \verb+hfilter+, which are both based on the Haar-Transform \citep{fritze77}. By testing these filters with different settings, they reported that while both are able to enhance faint features, neither is clearly an improvement over Gaussian filtering. A Gaussian filter is then preferred because it is computationally more efficient than other filters. By experimenting with the parameters of the Gaussian filter, they also reported that the best enhancement of faint extended features is achieved with a kernel scale of $\sigma = 7$. We have carried out our own tests and reached similar conclusions, finding that the best compromise between enhancing diffuse structures and preserving image resolution can be found at a $\sigma$ of $5$ to $7$ pixels, which at a distance of $20\ \mathrm{Mpc}$ corresponds to roughly $2$ to $3$ kpc. This works because the diffuse features of interest in this study have scales of a few kiloparsecs, therefore removing fluctuations on smaller scales makes them easier to detect by visual inspection. Convolutions with broader kernels take longer to compute without achieving higher detectability. Figure (\ref{fig:proc}) shows an example of the use of this enhancement technique to reveal the giant shell around NGC 4414. Whenever possible, we have added color insets of the central galaxies to the stretched images in order to visualize the relative extent of each galaxy and its low surface brightness halo.

%\LEt{because it occurs so frequently: when two words are a compound modifier to a third, they are hyphenated when there is danger of confusing which word belongs to which other word. For "low surface brightness", "low" belongs to "brightness", not to "surface", and it is clear that it wouldn't be "low surface" or "brightness halo", so there's no need to hyphenate for clarity. More than 2 words to modify a third can be awkward, so I rephrased previous instances, but this is just one option}

%\begin{figure*}
%\renewcommand*\thesubfigure{(A\arabic{subfigure})}s
%  \begin{center}
%    \includegraphics[width=\linewidth]{proc3.png}
%  \end{center}
%  \caption{NGC 4594, with a field of view of 22 $\times$ 38 arcminutes (north upwards). Left panel: SDSS color composite image from public archive, given as a reference. Right panel: Final product of our work ($I_f$) with a stretched greyscale overlaid with the original colormap, showing the extent of the stellar halo, and a conspicuous, looped stellar tidal stream south of the image. Right panel: image taken from the Stellar Tidal Stream Survey.}
% \label{fig:proc}
%\end{figure*}

\subsection{Photometric calibration and distribution of the surface brightness limit}
\label{sec:data_photCal}

An important issue for this work is to quantify the depth to which SDSS data allow us to explore faint stellar halo structures. In addition, the mean surface brightness limits of our images must have a narrow distribution to avoid image-to-image variance biasing any statistics we derive from visual inspection. Since the SDSS data were taken over a period of several years, we have to verify that the surface brightness limits of the images of different galaxies do not reflect variations in the quality of flat-fielding and the sky conditions during the observations, such as transparency and lunar phase (we note that SDSS generally observed in dark sky conditions; \citealt[e.g.,][]{Eisenstein11,york00,gunn06}). Scattered light due to bright stars in the vicinity of a galaxy will also contribute to fluctuations in depth. Large differences in seeing, depth, and the variance of depth across the image can lead to an important bias in the statistics of faint overdensities in our galaxy sample, since in some cases, non-detections of streams could be due to observational effects.

To quantify this, we measured the surface brightness limit of each image in our sample as follows. First, we performed a photometric calibration to the SDSS $r$ band for the coadded images, using the same approach as our previous studies of stellar tidal streams \citep{Chonis2011} and dwarf satellite galaxy populations \citep{javanmardi2016}. We chose the SDSS $r$ band to be consistent with other optical studies. All  297 processed images were calibrated using the semi-automatic pipeline developed (and successfully demonstrated) by the DGSAT\footnote{Dwarf Galaxy Survey with Amateur Telescopes} project \citep{javanmardi2016}.

Given the similarity between the effective bandpass of the stacked SDSS images and the wide-band luminance filter (used in the DGSAT), the calibration of our stacked images to the $r$ band requires a color correction, taking the form
\begin{equation}
  r_{cal}=c_0 \ell_{\rm{stacked}} +c_1 (g-r) + c_2,
\end{equation}
\noindent where $r_{cal}$ is the calibrated $r$ magnitude, $\ell_{\rm{stacked}}$ is the magnitude measured in the `pseudo-luminance' band of the stacked image, $c_0$ fixes the linear relation between these two magnitudes, $c_1$ corrects for a  color dependence, and $c_2$ is the magnitude zero-point correction. The constants $c_i$ are obtained using a set of calibrating stars in each image. These stars are selected using an automated statistical approach (rather than by hand), using the SDSS $g$ and $r$ band as standard magnitudes.

First, SExtractor \citep{sextractor96} was used to detect all the objects in each image. The detected objects were then cross-matched with the most recent SDSS photometric catalog, and only stars with $r\geq 15$, $0.08 < (r-i) < 0.5$ and $0.2 < (g-r) < 1.4$ passed on to the next step \citep[see ][]{chonis08}. At this point, very many stars are available for calibration of each image. The SDSS r-band magnitudes of the stars were compared iteratively to $r_{cal}$ calculated from each image and the $c_i$ for each stacked image obtained by a $\chi^2$ minimization.

\begin{figure}
  \begin{center}
    \includegraphics[width=\columnwidth]{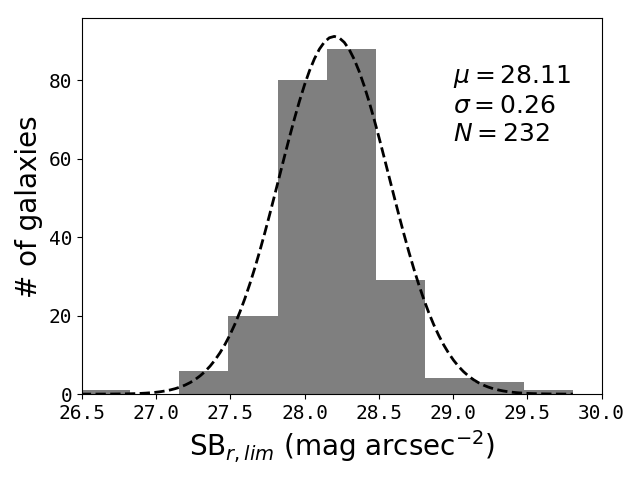}
  \end{center}
  \caption{Distribution of the surface brightness limit in our sample with a Gaussian fit, using the $232$ images in the sample with $N_\mathrm{star}>50$ stars after the $\sigma$ clipping as described in Sec. \ref{sec:data_photCal}. We obtained an average value of $28.11$, with a standard deviation of $0.26$.}
  \label{fig:hist_sblim}
\end{figure}

Next, any star with $\Delta r\equiv r_{SDSS}-r_{cal}$ deviating by more than twice the standard deviation $\sigma$ from the best-fit relation was discarded and the calibration relation was fit again to obtain new $c_i$. This clipping was repeated until no $2\sigma$ outlier remained, which gave us the final $c_i$ for each image. The standard deviation of $\Delta r$ provides an estimate of the uncertainty in the calibration and is added in quadrature when we report the uncertainty in magnitudes for each image. See \citet{javanmardi2016} for further details of this approach to photometric calibration.

After calibrating the data set to the SDSS $r$ band, the limiting surface brightnesses of our images were determined following the approach described in \citet{dmd10}. In short, to estimate the (residual) sky background, we measured the standard deviation in random sky apertures of $3$ arcseconds in diameter and computed the surface brightness corresponding to five times the standard deviation. Figure \ref{fig:hist_sblim} displays the distribution of the surface brightness limit of all images in our sample due to the mean sky background, showing that the data used in this work are sufficiently homogeneous in terms of quality and depth. We conclude that the mean sky surface brightness limit of our sample in the $r$ band is SB$_{r,lim} \approx 28.1\ \pm 0.3\ \mathrm{mag/arcsec}^2$. This means that for the purposes of the following analysis, we can neglect variations in depth as a significant source of bias in the statistics of low surface brightness features recovered by our visual inspection.

%\subsection{Surface brightness limit distribution}
\subsection{Deeper follow-up of tidal feature candidates}

  Although our processed SDSS images reach a surface brightness limit in the $r$ band (SB$_{r,lim}$) that would conventionally be considered `deep', they are still only deep enough to reveal the brightest structures (if any) in the halos of our target galaxies. In some cases, the low S/N detection of a particular feature and artifacts in the image significantly reduce our confidence in the nature of the detection and/or its interpretation as a signature of tidal disruption. Better (i.e., deeper) data are necessary to improve confidence in these detections.

\begin{figure}
\centering
\renewcommand*{\arraystretch}{0}
\begin{tabular}{*{2}{@{}c}@{}}

\begin{overpic}[width=0.5\linewidth]{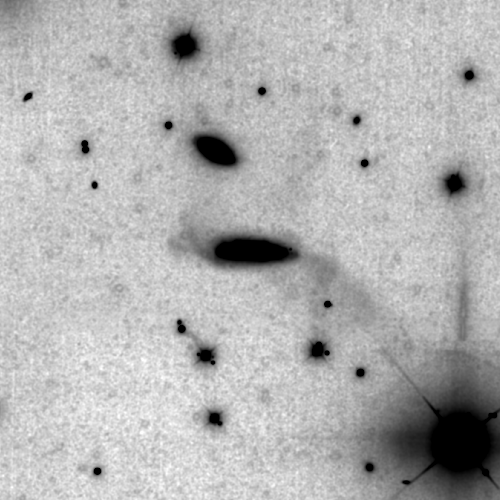}
\put(60,2){\colorbox{white}{\parbox{0.4\linewidth}{NGC 5506}}}
\end{overpic}&
\begin{overpic}[width=0.5\linewidth]{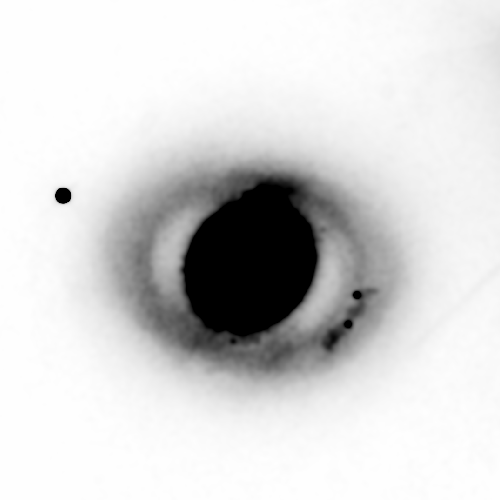}
\put(60,2){\colorbox{white}{\parbox{0.4\linewidth}{NGC 2859}}}
\end{overpic}\\
\begin{overpic}[width=0.5\linewidth]{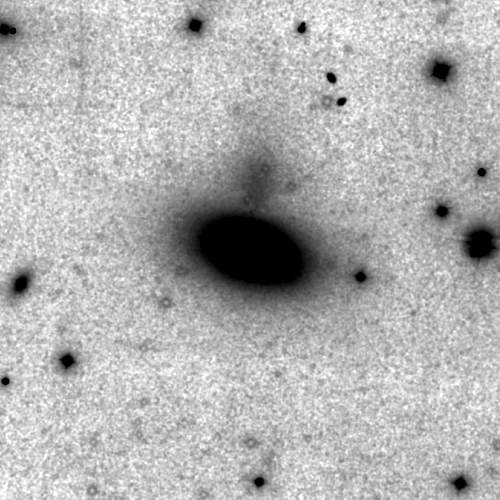}
\put(60,2){\colorbox{white}{\parbox{0.4\linewidth}{NGC 3489}}}
\end{overpic}&
\begin{overpic}[width=0.5\linewidth]{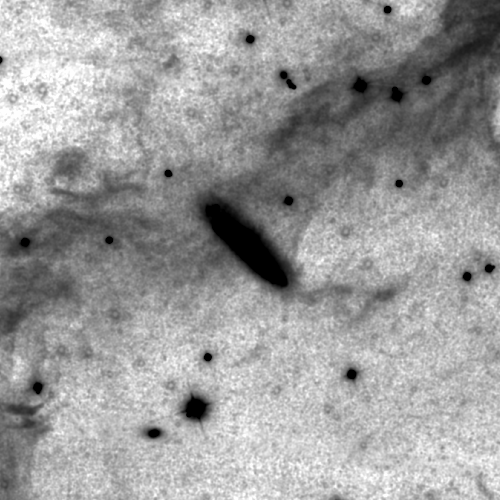}
\put(60,2){\colorbox{white}{\parbox{0.4\linewidth}{NGC 7497}}}
\end{overpic}\\

\end{tabular}
\caption{Examples of very faint diffuse overdensities of different types found during our analysis: i) giant stellar warps of a galactic disk (NGC 5506); ii) a stellar ring with two seemingly interacting, partially disrupted cores embedded in it (NGC 2859); iii) an image artifact resembling a giant satellite (NGC 3489); and iv) extensive Galactic cirrus around NGC 7497. All these images have been processed with our technique, adapted from \citet{miskolczi11}, using stacked SDSS $g$, $r,$ an $i$ band images, with a field of view of 30 arcminutes. North is up, east to the left.}
\label{fig:examples}
\end{figure}

  Figure \ref{fig:examples} shows some examples of different types of overdensities found in our search. These illustrate cases in which it is ambiguous whether well-detected overdensities are the result of minor mergers or perturbations of the central galaxy, such as extended stellar warps (e.g., NGC 5506), rings (e.g., NGC 2859) or other tidal distortions \citep[e.g.,][]{trujillo2009}. Significant sky background fluctuations or extended Galactic cirrus (e.g., NGC 7497) are also a well-known problem for the detection of tidal streams \citep[e.g.,][]{Duc15}. Finally, NGC 3489 is a clear example of a typical artifact in the
  SDSS images, a reflection from a bright star that resembles a diffuse satellite interacting with the central galaxy.

 % Top left panel shows NGC 5506 with two tidal structures rising northwards from the left side; and a huge, diffuse, amorphous cloud emerging towards the south. These features could be related to disk deformation due to tidal forces, and might not be stars accreted from minor merging events. Top right panel shows the presence of a ring and two very rare, seemingly interacting core-like structures with trailing and leading trails embedded in the disk. Assuming these cores actually belong to the halo of NGC 2859, it would be the very first case of tidal streams detected inside a detached halo ring. Bottom left panel shows a typical case of a confirmed image artifact emulating real diffuse features. Bottom right panel shows an extreme case of confirmed Galactic cirrus contaminating the sample, mimicking the diffuse profiles highlighted by the method described in this paper.

We have explored the availability of deep images for our stream candidates in two separate sources of additional optical data described below. Unfortunately, these additional surveys currently cover a smaller sky area and have fewer bandpasses than the SDSS, and hence are not suitable for the statistical analysis we attempt with SDSS data in this paper.

  %as shown in Figure \ref{fig:decam_final}. Table \ref{tab:confirmed} shows galaxies in our sample that have been previously reported and confirmed to present stellar tidal streams.

\subsubsection{Stellar Tidal Stream Survey}
\label{sec:data_stss}

\begin{figure*}
\renewcommand*\thesubfigure{(A\arabic{subfigure})}
  \begin{center}
   \includegraphics[width=\textwidth]{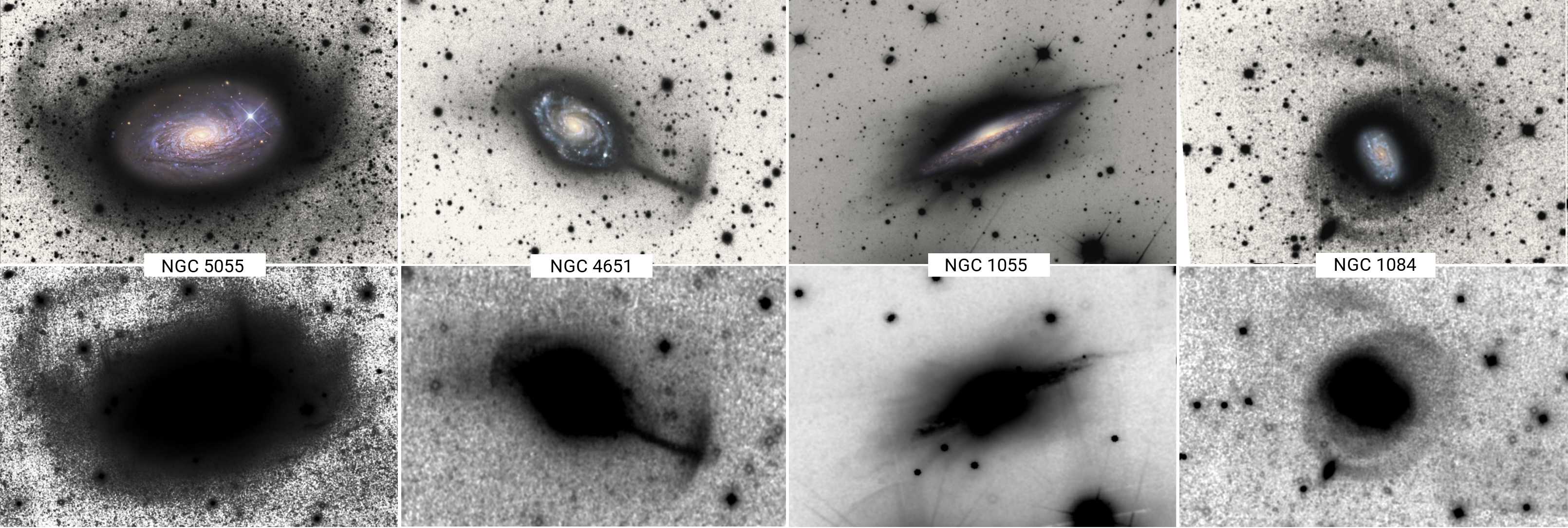}
  \end{center}
  \caption{Comparison between our processed SDSS images and the deep images from the STSS (see Sec. \ref{sec:data_stss}) for the spiral galaxies (from left to right) NGC 5055, NGC 4651, NGC 1055, and NGC 1084. Top row: Images taken from \citet{dmd10}. Bottom row: This work, using SDSS data, after processing as described in \citet{miskolczi11}, stacking $g-$, $r-,$ and $i$ -band images. Even at the shallower depth reached by the SDSS, high-confidence detections can be made with our technique. North is up, east to the left.}
  \label{fig:comp3}
\end{figure*}

Figure \ref{fig:comp3} shows a comparison of our results to the ultra-deep observations of the STSS \citep{dmd10} for a set of well-known diffuse features. As found by \citet{miskolczi11}, the filtering technique we used renders visible in SDSS images the majority of features reported so far by the robotic amateur telescope observations in the STSS pilot survey, although with a lower S/N because of the brighter surface brightness limit of the SDSS. The lower quality of our SDSS data compared to those of \citet{dmd10} is mainly explained by the short effective exposure times of the individual broadband SDSS observations. This complicates the classification of very faint overdensities, since the lower S/N makes it harder to distinguish actual tidal features from  overdensities related to  Galactic cirrus or image artifacts. Deeper follow-up observations are necessary to classify these ambiguous detections. %Overall, it can be seen that our approach reaches a surface-brightness regime in which the majority of the features detected in that survey can be recovered. However, the lower signal-to-noise of the SDSS images (mainly due to their short exposure time) makes it more difficult in some cases to characterize the overdensities and disentangle tidal streams from diffuse features with different origins (e.g., perturbations to the structure of the central galaxy, image artifacts, etc.; see Fig. \ref{fig:examples}).

\begin{figure*}
\begin{overpic}[width=\linewidth]{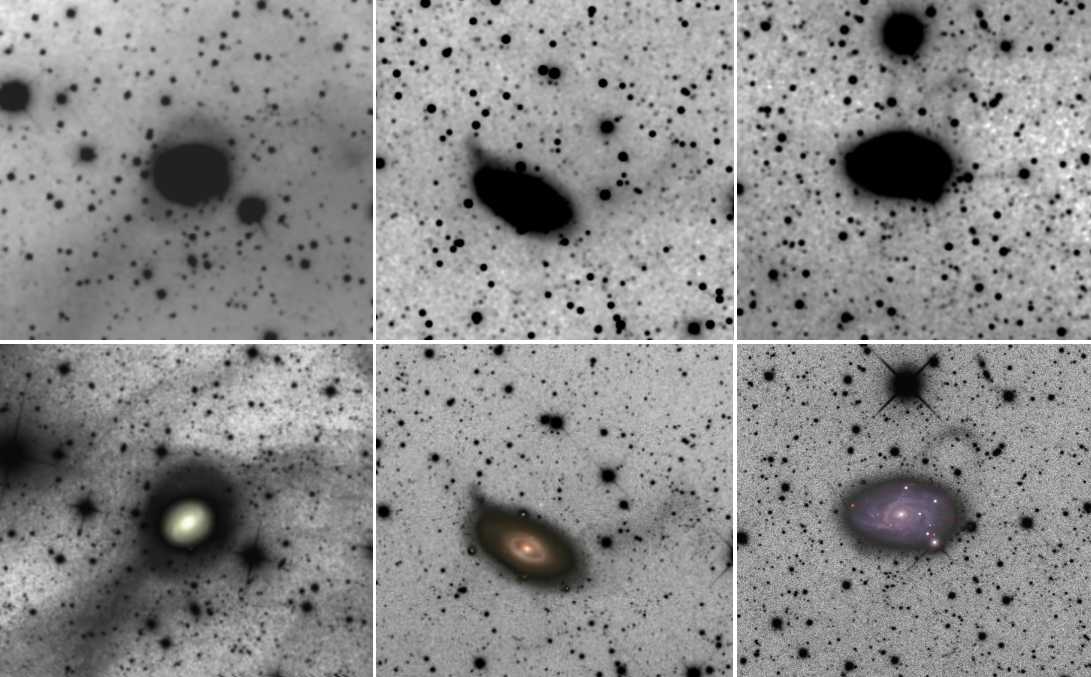}
\put(12,30){\colorbox{white}{NGC7743}}
\put(46,30){\colorbox{white}{NGC5750}}
\put(79,30){\colorbox{white}{NGC3041}}
\end{overpic}
\caption{Examples of some follow-up images used to confirm stellar tidal stream candidates in our sample. The top row shows the SDSS $g$-$r$-$i$ stacked images for NGC 7743, NGC 5750, and NGC 3041 (see Sec. \ref{sec:res}) processed as described in Sec. \ref{sec:data}, and kindly provided by A. Miskolczi. The bottom row shows the deep images obtained by the STSS (Mart\'inez-Delgado et al., in prep.) for the same objects, but with a luminance filter. In all cases, the deeper images detect additional features or reveal a more detailed morphology of the features detected in the SDSS images.}
\label{fig:misk_vs_stss}
\end{figure*}

Figure \ref{fig:misk_vs_stss} illustrates the benefits of deeper follow-up with observations like those of the STSS for three galaxies in our sample. The first column presents NGC 7743, a galaxy with an apparently clear stellar tidal stream candidate in its outskirts; however, with STSS observations this feature is shown to be (at least predominantly) Galactic cirrus. The second column presents NGC 5750, which shows a remarkable irregular substructure apparently emerging from its disk; deeper STSS observations reveal an additional overdensity on the other side of the galaxy, which is only barely visible at the detection limit of our SDSS images. Finally, the third column presents NGC 3041, in which an arc-like feature is apparent to the north of the galaxy. The amateur telescope data strongly support the interpretation of this feture as a great-circle stream.

\subsubsection{Dark Energy Camera Legacy Survey}

We have also searched for images of galaxies with visible low
surface brightness features (Table \ref{tab:main_table}) in the optical images from the third data release (DR3) of the Dark Energy Camera Legacy Survey \citep[DECaLS, ][]{decals16}. This survey uses the Dark Energy Camera (DECam), a wide-field CCD imager at the CTIO Blanco 4 m telescope, to obtain optical imaging covering 14,000~deg$^2$ in three optical bands ($g$,$r$,$z$). Since the footprint is mostly in the equatorial and southern sky and only a fraction of the DECaLS data have been publicly released, only a few targets have been imaged with sufficient quality and depth to aid in the interpretation of our SDSS images. In the publicly available DECaLS data, we found that three of our targets were imaged in the $g$, $r$, and $z$ bands, confirming the detected streams or diffuse-light substructure in the halo from our analysis. Although some sky regions have been imaged in just one DECaLS band so far, we are able to improve our confidence in some low surface brightness features with even these data (see Sec. \ref{sec:res_confirmed}). Regarding background subtraction around large objects, it must also be noted that the DECaLS data reduction for large sources has not yet been optimized to the same extent as in the SDSS \citep{Blanton2011}. This explains the rectangular patches in DECaLS images with poor subtraction around large galaxies, which sometimes mimic diffuse galactic structure.

%\subsection{Comparison with previous work}
%\label{sec:comp}

 %\citet{miskolczi11} calibrated images of galaxies taken from the SDSS archive, %processed them in an automated manner, and visually inspected them for possible %tidal streams.

%Figure \ref{fig:hist_sblim} displays the distribution of the surface brightness limit of all images in our sample, showing that the data used in this work are sufficiently homogeneous in terms of quality and depth. This means that, for the purposes of the following analysis, we can neglect variations in depth as a significant source of bias in the identification of low-surface-brightness features by visual inspection. We conclude that the mean surface brightness limit of our sample in the $r-$band is SB$_{r,lim} \approx 28.1\ \pm 0.3\ \mathrm{mag/arcsec}^2$.

%This means that we can safely assume that our data do not present much variation in terms of how they were collected and processed by the SDSS pipeline, which is important in any systematic search in order to keep the detectability reliable, since different conditions during the imaging stage would add a systematic effect or bias to the sample. This effect can also be understood by recalling the studies from \citet{johnston08}, where the ability to detect streams over the background level greatly depends on the surface brightness limit of the image.

\section{Results}
\label{sec:res}

\subsection{Confidence of detections and morphologies}
\label{sec:res_dcl-morph}
For all of the galaxies listed in Table \ref{tab:main_table}, we estimated our confidence in the detection of faint tidal features on a five-point scale, similar to the scheme used by \citet{Atkinson2013}. We refer to it as the detection confidence level (DCL), reflecting our certainty that a tidal feature with
low surface brightness that is associated with the target galaxy was detected in an image, as follows:

\begin{description}
\item[0:]No detection of any sort, or high confidence that any candidates are perturbations of the central object (spiral arms, cirrus,
etc.).
\item[1:]Very uncertain detection of a feature at a S/N too low to judge either the quality of the detection or its tidal nature.
\item[2:]Possible detection of a low surface brightness feature, but with low confidence in a tidal origin ($\sim 50\%$ certain).
\item[3:]Strong detection judged highly likely to be of tidal origin, but without support from any data other than our own.
\item[4:]Strong detection where a tidal origin is supported by other data.
\end{description}

In the same table, we also provide an approximate visual classification system for the morphology of the most common features we detect using the following categories, which are not mutually exclusive:

\begin{description}
\item[S] for classic shells. Disconnected, coherent arcs of material usually concentric with the central galaxy.
\item[C] for any coherent, curvilinear features seen in the image (that are not shells). This includes arcs, plumes of debris, and looped structures of gas or stars surrounding their host.
\item[Sph] for spheroids and diffuse satellites in the process of disruption, suggestive of very low surface brightness galaxies.
\item[E] for extensions of the central galaxy, including but not limited to warps and spiral arms, and some unclassifiable overdensities clearly connected to the disk.
\item[O] for any other less common type of features not included above, but suggestive of tidal interactions: fuzzy clouds, spikes, wedges, irregular filaments, etc.
\end{description}

These categories are intended as an simple indication of the appearance of the features we detect. Stronger conclusions about the true physical nature of these features are beyond the scope of this work, and in most cases would require support from techniques other than photometry.

\subsection{Sample statistics}

By visual inspection of the processed images, we determine that $51$ of the $297$ galaxies in our sample show either clear or potential signatures of diffuse overdensities in their outskirts above our surface brightness limit ($28.1\ \mathrm{mag~arcsec^{-2}}$). Table \ref{tab:main_table} describe these galaxies and their associated low-surface-brightness features. Of these $51$ targets, $28$ show overdensities that we judge to be either stellar or gaseous tidal features on the basis of other, deeper observations. A further $23$ objects for which we currently lack deeper observations show overdensities that are likely tidal feature candidates (listed in Table \ref{tab:main_table} as features with DCL 1 and 2; see Sec. \ref{sec:res_dcl-morph}). Hence a conservative estimate for the frequency with which such features occur in our volume- and mass-limited sample of the local Universe is $\approx 9\%$. This would rise to $\approx 17\%$ if all candidates were confirmed by deeper follow-up observations. Considering only previously published features together with the new discoveries reported in this paper, we estimate that $\approx 14\%$ of galaxies in the local Universe exhibit diffuse tidal features brighter than $28.1\ \mathrm{mag~arcsec^{-2}}$ in the SDSS $r$ band. As a reference, all galaxies in our sample that do not show any evidence of diffuse-light structures are listed in Table \ref{tab:negatives}.

\begin{figure*}
\includegraphics[width=\linewidth]{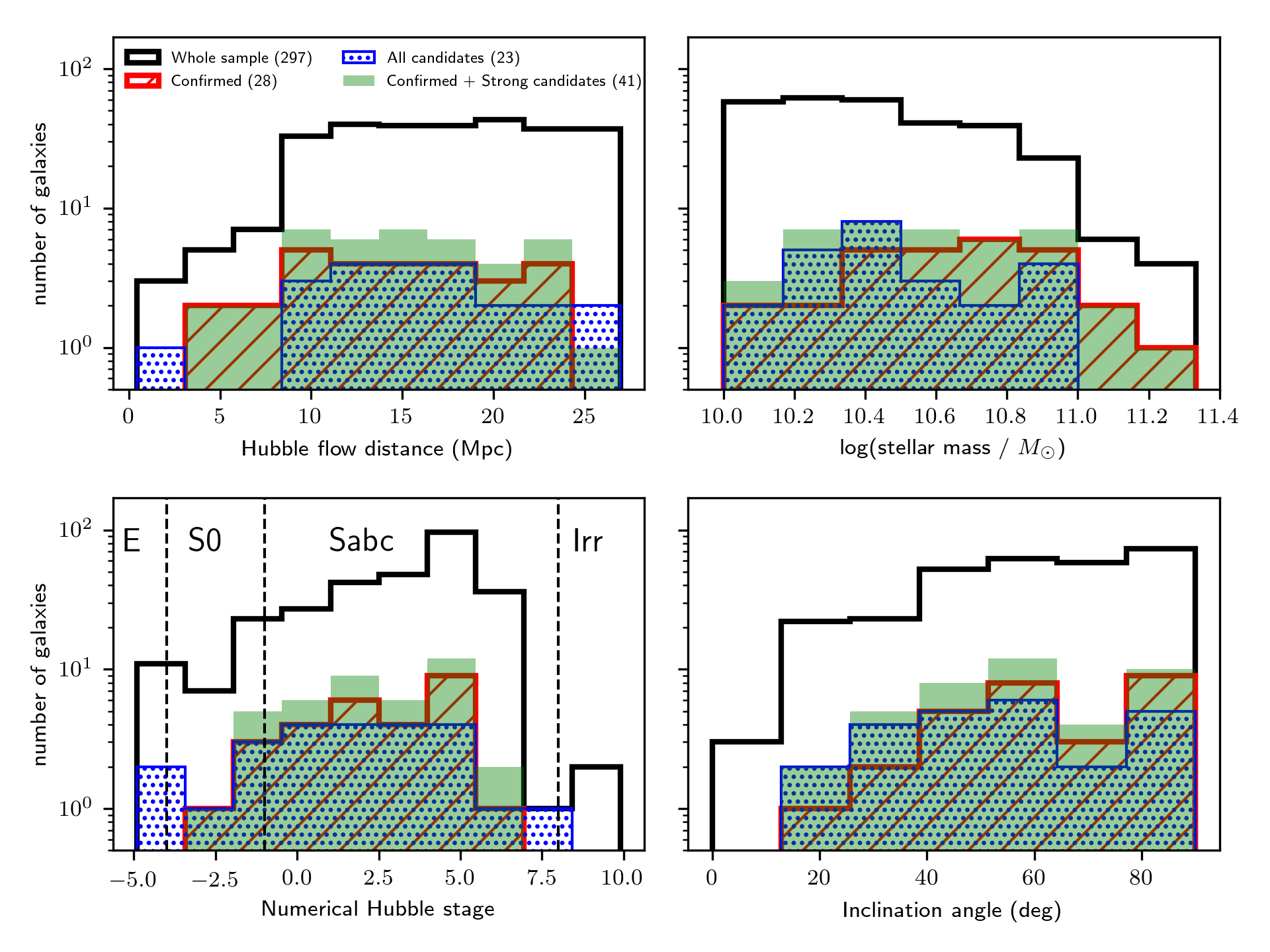}
\caption{Histograms of the diffuse-light features found in our whole sample, with or without overdensities, as a function of target distance, stellar mass, morphology and inclination angle. The distribution of all 297 galaxies in our sample is shown in black, while histograms in color correspond to the galaxies listed in Table \ref{tab:main_table}, our main results. The red distributions represent 28 confirmed features, previously known and new (with a DCL of 4; see Sec. \ref{sec:res_dcl-morph}). Unconfirmed feature candidates (with a DCL of 1 and 2) are represented by blue dotted histograms. Confirmed streams and strong candidates (i.e., every feature with a DLC of 3 or 4) are grouped together in the solid green histogram ($41$ targets). For our limiting sky surface brightness of $28.1\ \pm 0.3\ \mathrm{mag~arcsec^{-2}}$, this implies that $\approx 14\%$ of the galaxies in the mass and volume limits our parent sample have detectable stellar overdensities in their outskirts. No significant biases are apparent in our sample of galaxies with diffuse overdensities (with respect to the S4G parent sample, black solid line) except for a somewhat flatter distribution of stellar mass and the lack of overdensities for galaxies more distant than $35$~Mpc.}
\label{fig:fractions}
\end{figure*}

Figure \ref{fig:fractions} shows histograms of our galaxy samples, including the parent sample and subsets of targets that show evidence of tidal streams\footnote{Instead of using the mean redshift-independent distances from NED, we here used the Hubble flow distances, $cz/H_0$, where $cz$ is the recessional velocity, and $H_0=75\ \mathrm{km/s/Mpc}$.}. Furthermore, it shows that above the surface brightness limit of our sample there are no significant selection effects arising from either morphology or inclination angle, consistent with a roughly isotropic distribution of features. There are some hints of an excess of low surface
brightness features at short distance and high stellar mass. The latter is compatible with the expectation of $\Lambda$CDM, where more massive galaxies are hosted by more massive DM haloes and are therefore more likely to accrete brighter satellites. Any selection effect with distance is likely to reflect the balance between this effect (large volumes include more bright galaxies) and the increasing difficulty of detecting low surface brightness substructures at larger distances.

We also computed a two-sample Kolmogorov-Smirnov (KS) test to provide a quantitative comparison between the distance, mass, morphology, and inclination angle distributions of the $297$ galaxies in our parent sample and those of the hosts of high-confidence tidal feature candidates from Table \ref{tab:main_table} (this means a DCL equal to 3 or 4, i.e., a comparison of the black and solid green histograms in Fig. \ref{fig:fractions}). The $p$-values obtained are shown in Table \ref{tab:pvalues}. At a significance of $p=0.05$, on the basis of any one of these four distributions, we cannot reject the null hypothesis that our sample of galaxies with overdensities is drawn from the same underlying population as the parent sample we selected from \sg. In other words, random sampling from our parent sample has a high probability of yielding distributions similar to those of our sample of galaxies with overdensities.

\begin{table}
\centering
\begin{tabular}{cc}

Variable     & $p$-value \\
\hline
Distance     & 0.113\\
Stellar Mass & 0.067\\
Hubble Stage & 0.241\\
Inclination  & 0.654

\end{tabular}
\caption{$p$-values obtained from a two-sample KS test applied to the histograms depicted in Fig. \ref{fig:fractions}; specifically, the distribution of confirmed tidal streams plus Class I candidates versus the whole sample of $297$ galaxies. We cannot reject the hypothesis that the distributions of the two samples are the same.}
\label{tab:pvalues}
\end{table}

\subsection{Confirmed stellar structures with low surface brightness }
\label{sec:res_confirmed}
\begin{figure*}
\centering
\renewcommand*{\arraystretch}{0}
\begin{tabular}{*{3}{@{}c}@{}}

\begin{overpic}[width=0.3\linewidth]{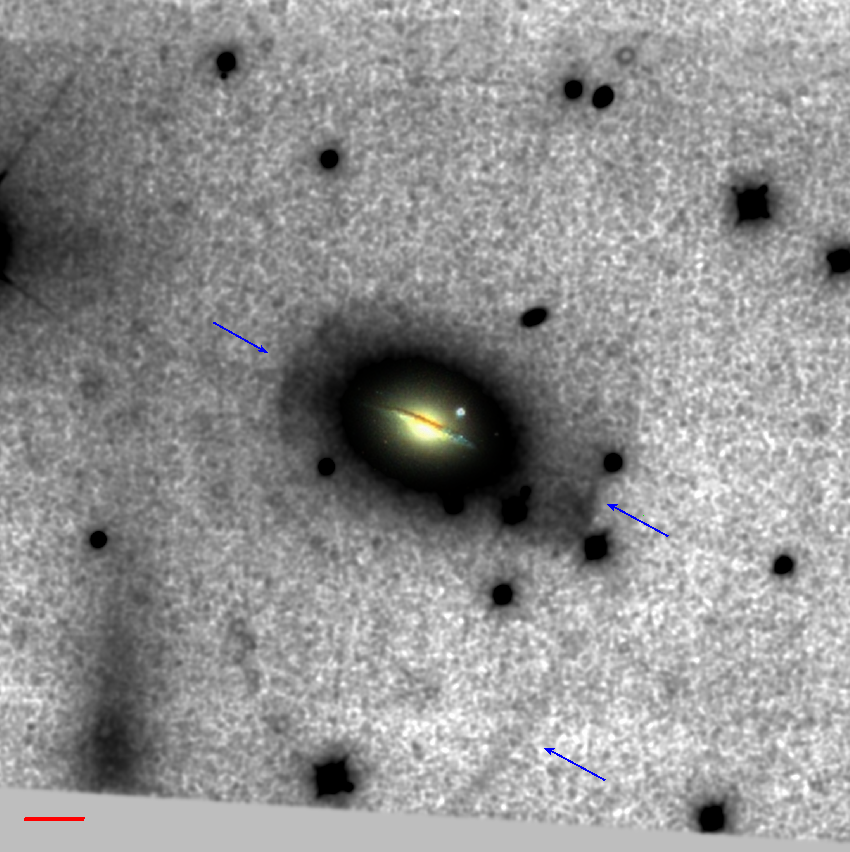}
\put(68,2){\colorbox{white}{\parbox{0.4\linewidth}{NGC 681}}}
\end{overpic}&
\begin{overpic}[width=0.3\linewidth]{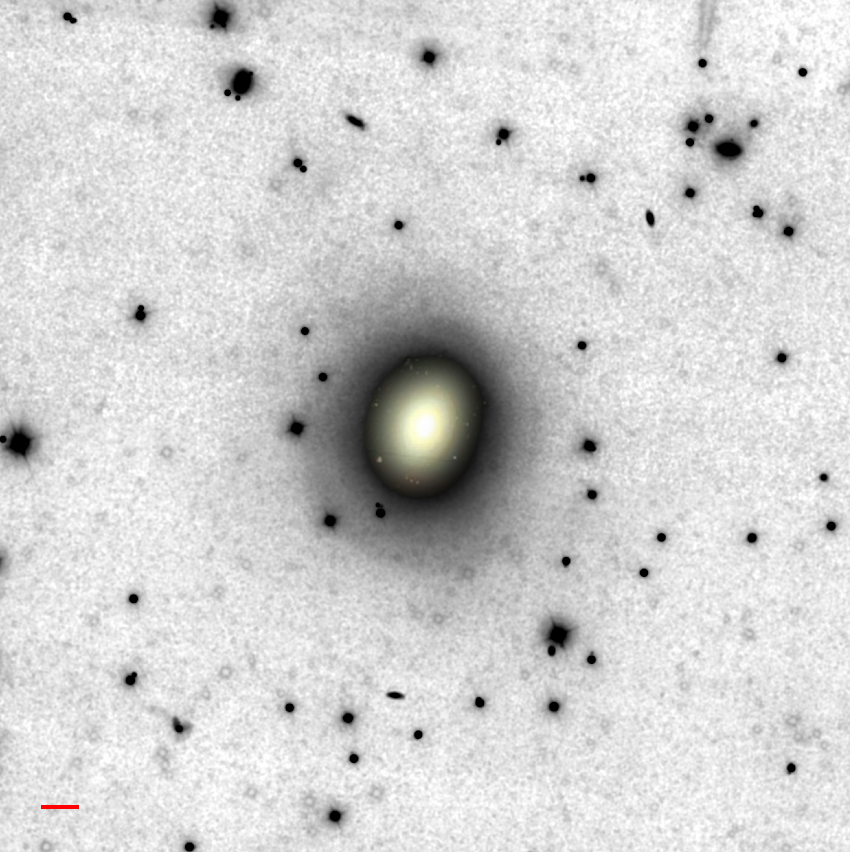}
\put(68,2){\colorbox{white}{\parbox{0.4\linewidth}{NGC 2775}}}
\end{overpic}&
\begin{overpic}[width=0.3\linewidth]{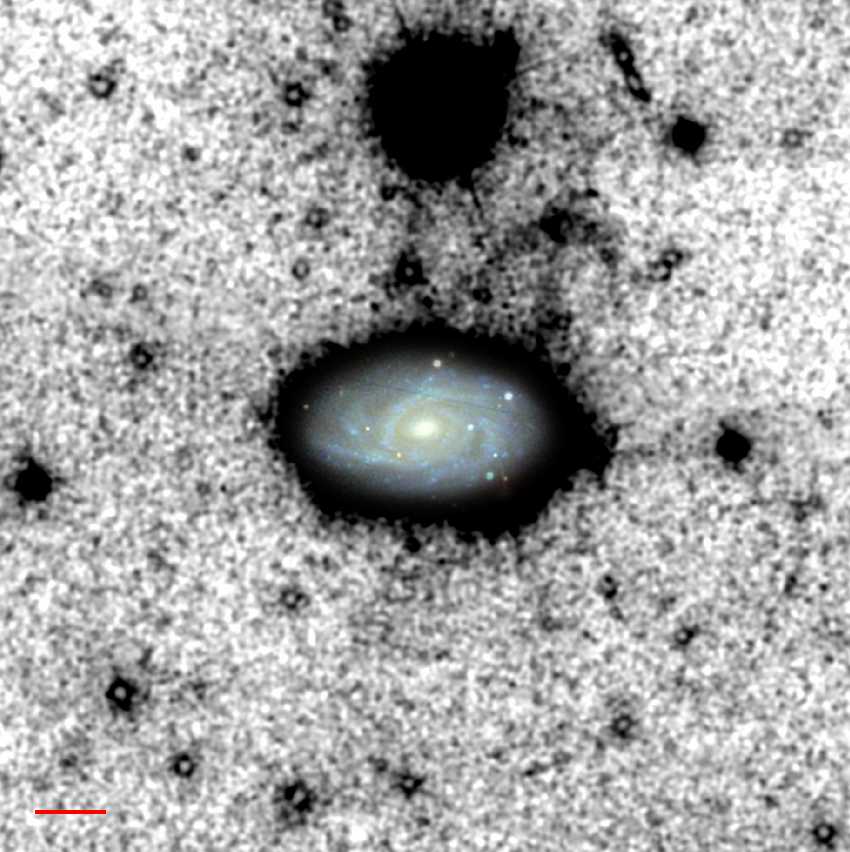}
\put(68,2){\colorbox{white}{\parbox{0.4\linewidth}{NGC 3041}}}
\end{overpic}\\

\begin{overpic}[width=0.3\linewidth]{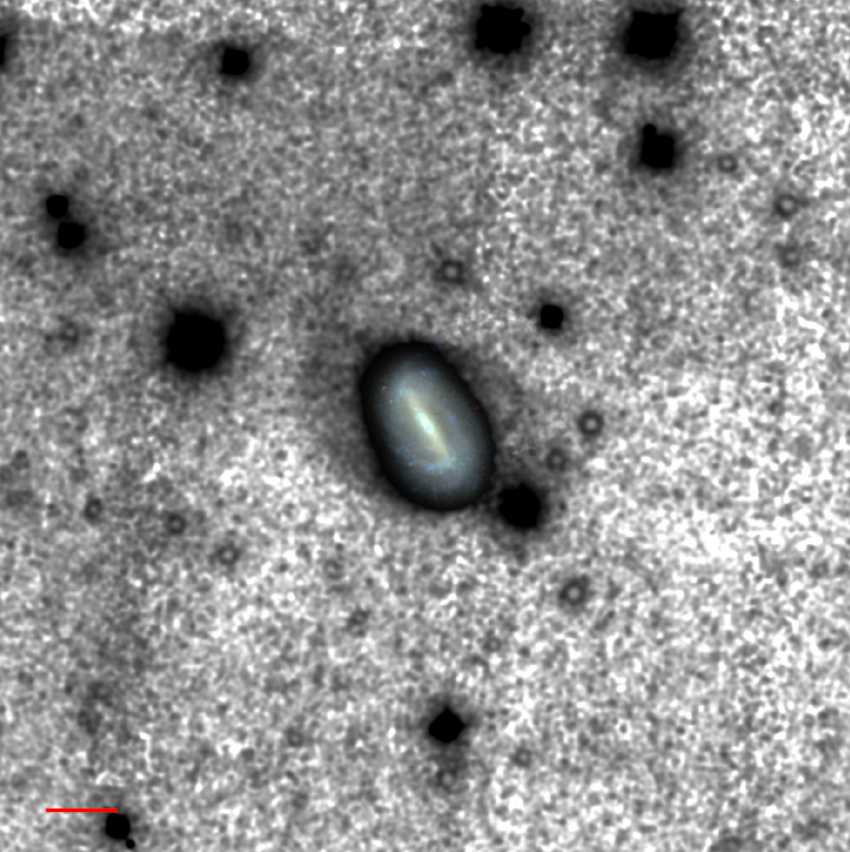}
\put(68,2){\colorbox{white}{\parbox{0.4\linewidth}{NGC 3049}}}
\end{overpic}&

\begin{overpic}[width=0.3\linewidth]{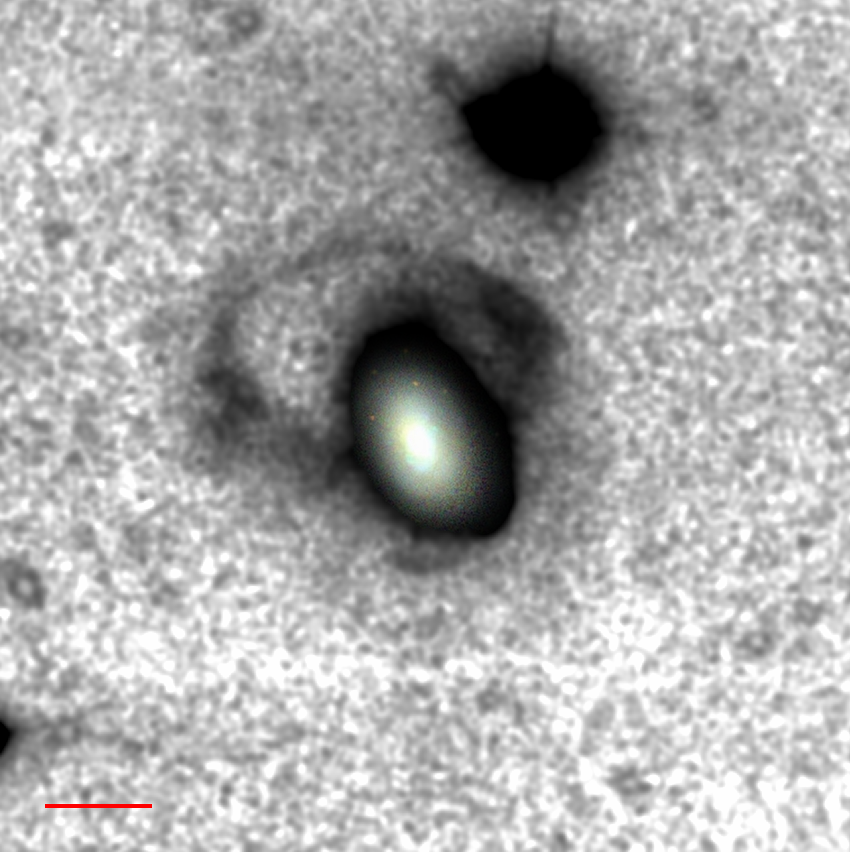}
\put(68,2){\colorbox{white}{\parbox{0.4\linewidth}{NGC 3611}}}
\end{overpic}&
\begin{overpic}[width=0.3\linewidth]{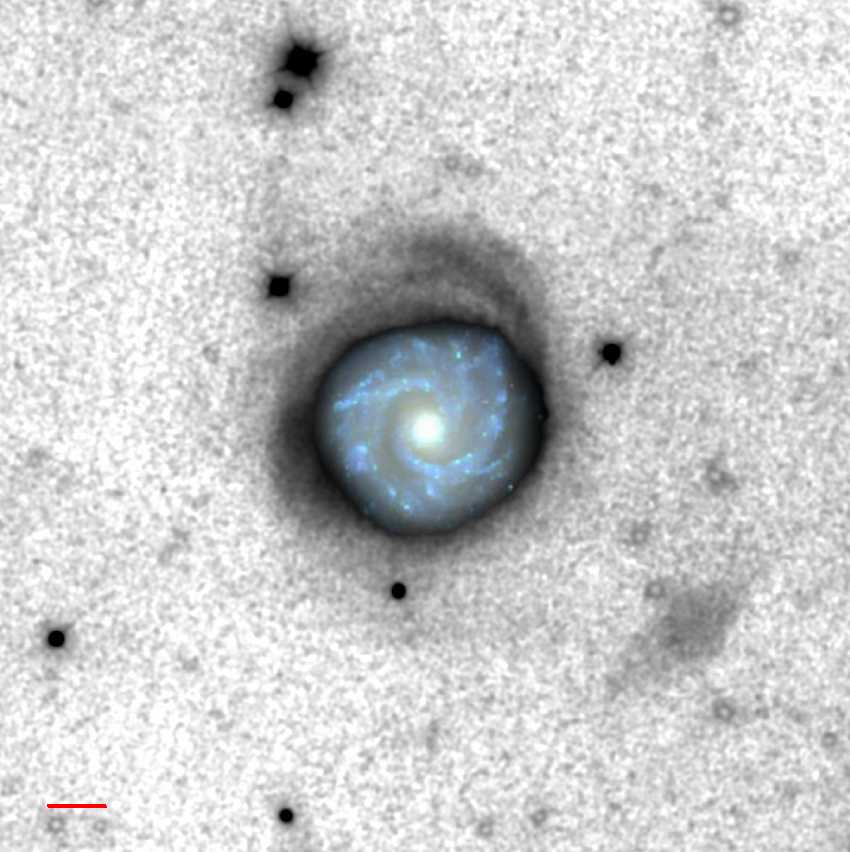}
\put(68,2){\colorbox{white}{\parbox{0.4\linewidth}{NGC 3631}}}
\end{overpic}\\
\begin{overpic}[width=0.3\linewidth]{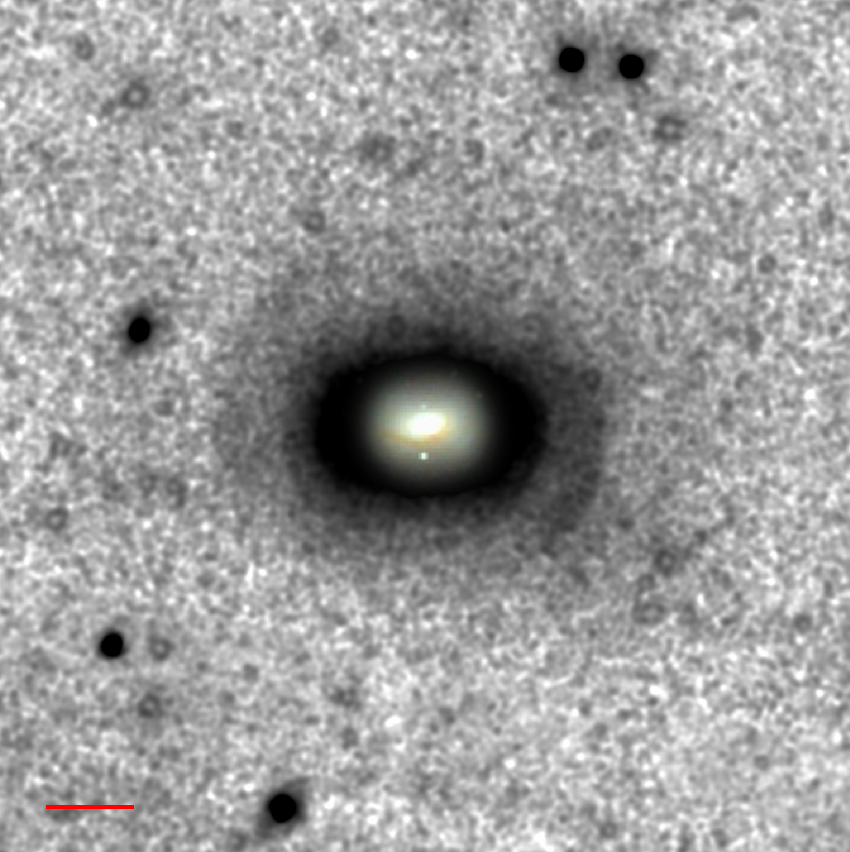}
\put(68,2){\colorbox{white}{\parbox{0.4\linewidth}{NGC 3682}}}
\end{overpic}&
\begin{overpic}[width=0.3\linewidth]{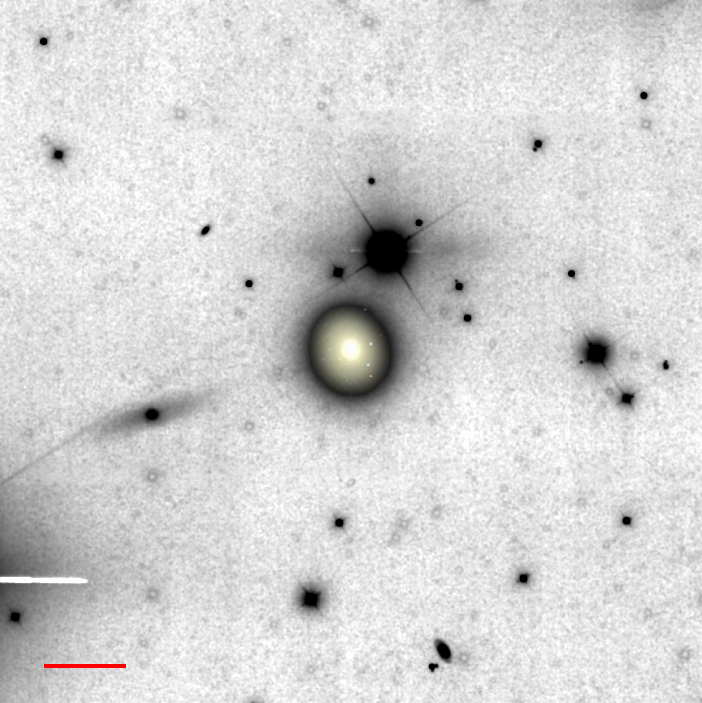}
\put(68,2){\colorbox{white}{\parbox{0.4\linewidth}{NGC 4203}}}
\end{overpic}&
%\begin{overpic}[width=0.3\linewidth]{ngc4631_paperins.png}
%\put(68,2){\colorbox{white}{\parbox{0.4\linewidth}{NGC 4631}}}
%\end{overpic}&

\begin{overpic}[width=0.3\linewidth]{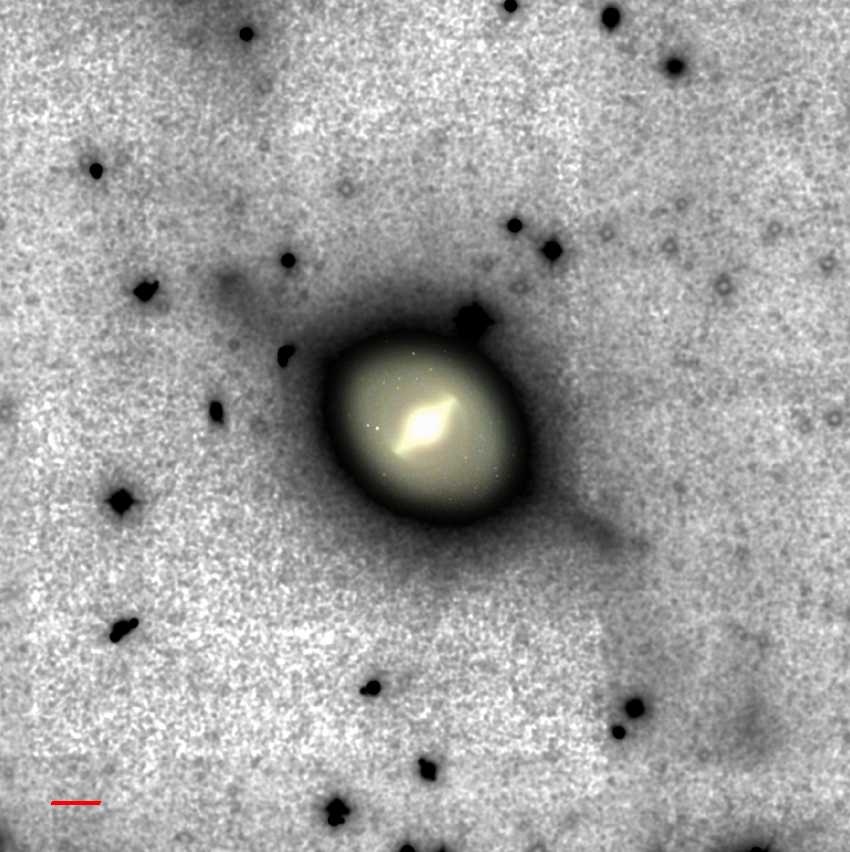}
\put(68,2){\colorbox{white}{\parbox{0.4\linewidth}{NGC 4643}}}
\end{overpic}\\
\begin{overpic}[width=0.3\linewidth]{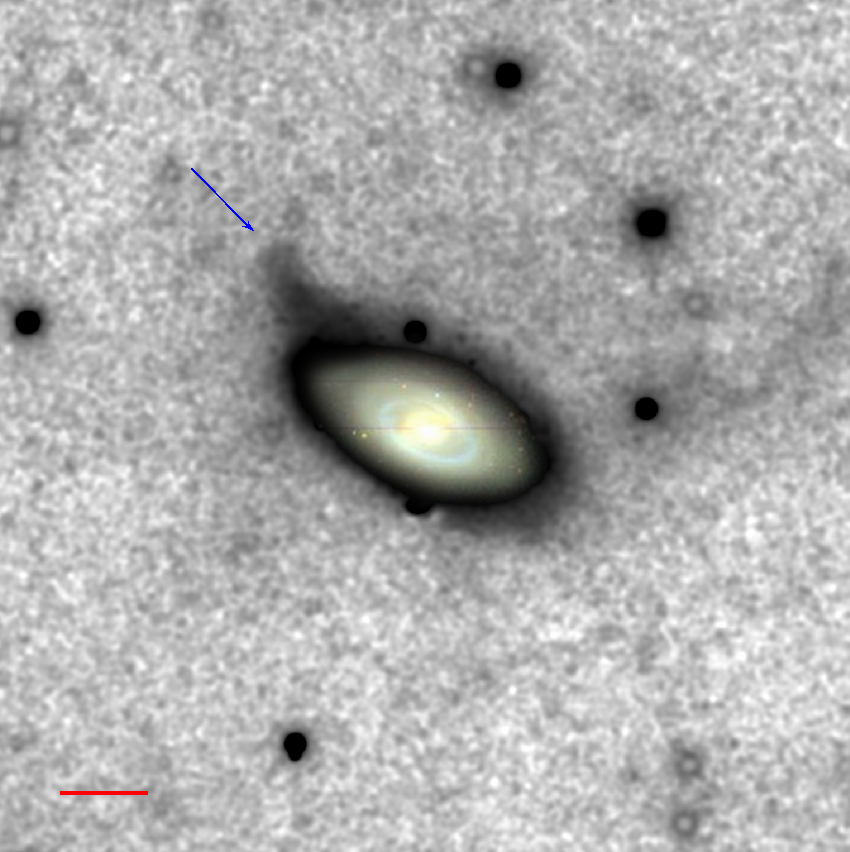}
\put(68,2){\colorbox{white}{\parbox{0.4\linewidth}{NGC 5750}}}
\end{overpic}&
\begin{overpic}[width=0.3\linewidth]{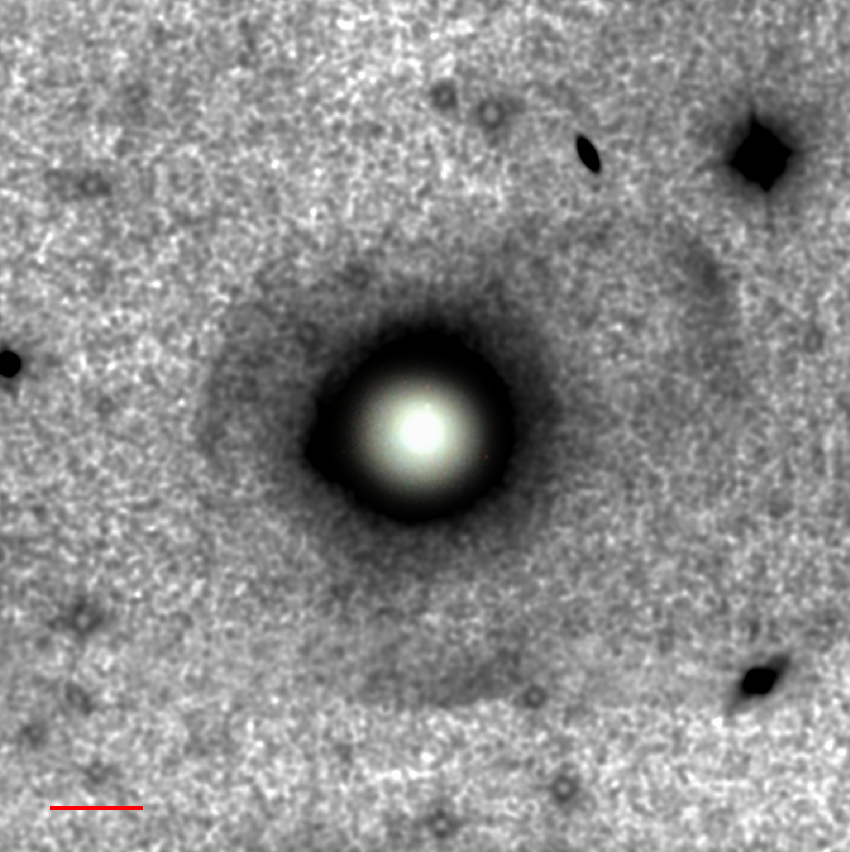}
\put(68,2){\colorbox{white}{\parbox{0.4\linewidth}{NGC 7742}}}
\end{overpic}&
\begin{overpic}[width=0.3\linewidth]{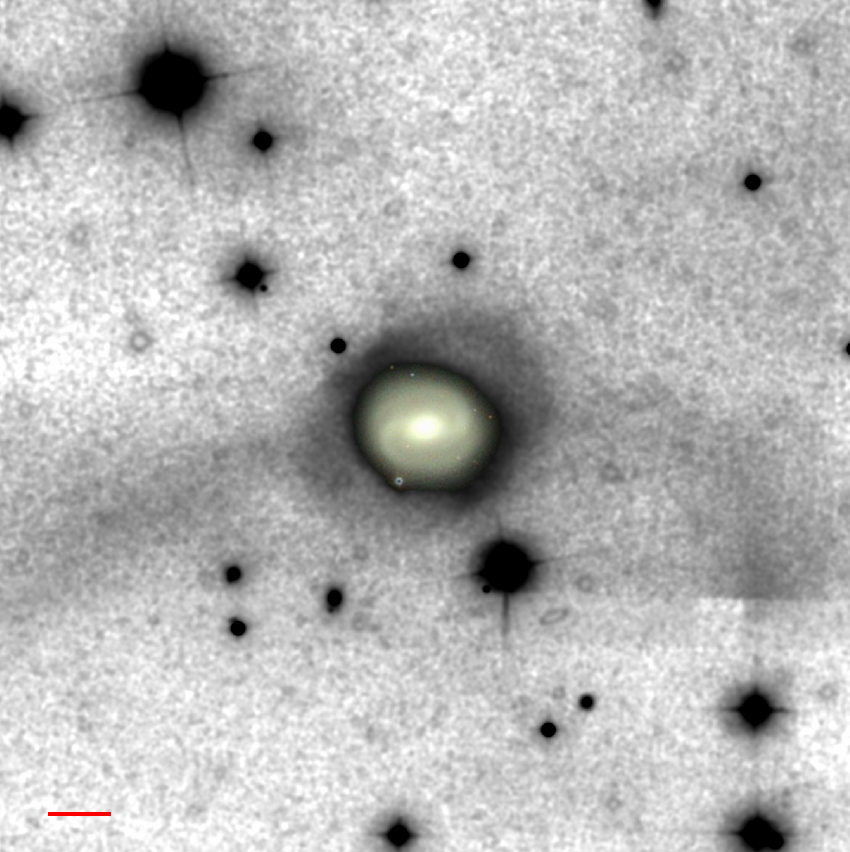}
\put(68,2){\colorbox{white}{\parbox{0.4\linewidth}{NGC 7743}}}
\end{overpic}\\

\end{tabular}
\caption{Confirmed tidal streams from Table \ref{tab:main_table} detected by stacking $g-r-i$ SDSS bands as described in this work. Streams already reported in previous publications are not included. The red lines indicate a scale of 3 arcminutes. In some cases, blue arrows indicate structures of interest described in Section \ref{sec:res_confirmed}. North is up and east to the left.}
\label{fig:confirmed}
\end{figure*}

Table \ref{tab:main_table} lists, among others, the $28$ tidal streams that we have confirmed. This list contains $12$ unpublished detections, including those found recently in the STSS (Martinez-Delgado et al., in preparation). Figure \ref{fig:confirmed} shows the corresponding images for each of their $12$ host galaxies. In these images, the disks of galaxies tend to dominate the field of view because the images have been significantly contrast-stretched to render the low surface brightness structures visible. These new streams are briefly described below. The estimated physical extent of these substructures has been calculated assuming the distance to the target taken from the NASA/IPAC Extragalactic Database (NED). It must be noted that the NED uses the mean value of redshift-independent distances. When no mean distance is reported, the Hubble flow distance is used instead.

\begin{description}
\item[NGC 681] is an edge-on disk galaxy with a prominent spheroid surrounded by two clear shells along the major axis, extending to $R=25\ \mathrm{kpc}$ southwest and $R=39\ \mathrm{kpc}$ northeast. Other possible arc-like features (concentric around the galaxy $R=10\ \mathrm{kpc}$ to the south and extending northeast from the south edge of image) are not apparent in DECaLS images (see Fig. \ref{fig:decam_final}).

\item[NGC 2775] is an unbarred spiral galaxy showing a prominent $\sim 29\ \mathrm{kpc}$ cloudy structure in its halo, reminiscent of a classical shell from \citet{dmd10}.

\item[NGC 3041] shows an arc-like stream with an extent along its longest dimension of $\sim 4\ \mathrm{kpc}$, northeast of the central galaxy. STSS and DECaLS data (Fig. \ref{fig:misk_vs_stss}) show this feature clearly, but they do not reveal any further detail. This may be the brighter part of a great circle structure similar to the Milky Way Sagittarius stream, but no surviving progenitor is apparent.

\item[NGC 3049] shows an arc-like feature east of the central galaxy, with a size of $\sim 3\ \mathrm{kpc}$. Another very diffuse substructure can be identified to the west, suggestive of a shell formed by tidal disruption. More definitive statements require deeper observations.

\item[NGC 3611] is well known for the peculiar $\sim 30\ \kpc$ bright off-center ring-like structure previously noted by \citet{Schweizer1990}. These authors favored merging as the origin of this feature, excluding the possibility of a disturbed polar ring. %The deepest view of this structure %provided by the follow-up DECam images
DECaLS data (Fig. \ref{fig:decam_final}) show two distinct features: a clear umbrella-like stream with shells on both sides of the galaxy (the most prominent to the east), and an incomplete blue ring or arc encircling the disk. The colors of both structures are clearly different, and it is unclear whether they have a common origin (e.g., the tidal disruption of a Magellanic-type dwarf galaxy).

\item[NGC 3631] shows a giant cloud at a galactocentric distance of $\sim 19\ \kpc$. This is very similar to the M83 stream \citep{Malin97}. It is not clear whether the structure is part of a very faint outer disk or a tidal structure in the galactic halo. This overdensity has also been observed by the STSS (Martinez-Delgado et al., in preparation), indicating that it is not an artifact in the SDSS image.

\item[NGC 3682] shows two classical shells on both sides of the central galaxy, with diameters of $\sim 2\ \mathrm{kpc}$.

%\item[\textbf{NGC 4041] on the other hand, presents a filamentary stream of stellar material of $\sim 15\ \mathrm{kpc}$ south of the target, but due to the orientation of the target it is not clear whether it is an extension of the disk, or a true, tidal structure extending beyond the halo.

\item[NGC 4203] shows a bright, partially disrupted and nucleated satellite southwest of the galaxy, with both a leading and a trailing tail of total length $\sim 13\ \mathrm{kpc}$.

\item[NGC 4569] is a spiral galaxy with a dIrr satellite (IC 3583) to the north, with an apparent interaction between the two\footnote{Although gravitational interaction has been ruled out by some authors \citep[e.g.,][]{Boselli16}.}. There is evidence of a shell-like overdensity on the northern side of the galaxy, although we cannot reject the possibility that this is an extended warp of the stellar disk.

%\item[\textbf{NGC 4519] is another good example of two short, $\sim 16\ \mathrm{kpc}$ filaments of debris emerging from the southern edge halo. Again, due to the orientation of the host galaxy, it is difficult to assess its physical origin.

\item[NGC 4643] shows a clear stellar tidal stream apparently perpendicular to the plane of the galaxy. DECaLS data show evidence for a progenitor in the northern tail. Assuming that both structures apparent in the image are part of the same feature (for example, an arc viewed edge-on), this feature has an extent of $\sim 73\ \mathrm{kpc}$. \citet{Whitmore90} reported an inner, edge-on arc structure in the main body of the galaxy that is also visible in our images, but not related to the giant tidal structure we report here.

\item[NGC 5750]  Both images from our analysis and the STSS deeper images (see Fig. \ref{fig:misk_vs_stss}) show a truncated overdensity west of the central galaxy, which resembles a faint, distorted satellite galaxy. In addition, an elongated, irregular feature east of the disk (clearly visible in the STSS image) could be part of a tidal stream associated with that satellite.

\item[NGC 7742] is a face-on unbarred Seyfert spiral galaxy, which shows three very distinct stellar arcs, possibly sections of a shell (or shells), each $16-17\ \kpc$ in diameter.

\item[NGC 7743] is a barred Seyfert spiral galaxy showing a giant, $18\ \kpc$ loop structure to the northeast. Galactic dust clouds dominate the field of view in longer exposures, as shown in the first column of Fig. \ref{fig:misk_vs_stss}.

%\item[\textbf{NGC 4762] is another interesting case of what seems to be the effects of disk warping coupled with mixed-type stream features, with approximate sizes of $\sim 18\ \mathrm{kpc}$ each. For NGC 4762 we report streams of different morphologies also thought to be present in DECaLS data...

%\item[\textbf{NGC 5506] displays a two arcs extending from the west and a big stellar cloud to the east, with apparent sizes of $\sim 5-6\ \mathrm{kpc}$ each.
\end{description}

 % In the case of NGC 3611, Schweizer \& Seitzer (1990) reported the discovery of a peculiar, off-centered ring of $30\ \mathrm{kpc}$ in diameter around this galaxy, discussing four hypotheses for its origin: (1) that is a background or foreground object (which the authors deemed rather unlikely); (2) that the ring is related to polar rings, but stating that there is no known mechanism for rings to become off-centered; (3) an actual tidal tail of an accreted companion, and (4) a peculiar case of another classical ring galaxy, including the disruption of a dwarf satellite. Any of these last two cases corresponds to the outcome of minor merging events shortly after the formation of the galaxy, which leads us to judge this case as another positive detection of diffuse light substructure also confirmed by DECaLS.

 %  NGC 5364 and NGC 4691 show nearly face-on galaxies with some evidence of either big clouds of gas and stars bound to its host that also have been reported here. For NGC 4762 we report streams of different morphologies also thought to be present in DECaLS data, while NGC 5740 shows what appears to be a truncated disk arm, but also possibly with tidal origin. In this case, the deeper data from DECaLS allowed us to better judge the nature of this arc, which is not as clear from our data analysis.

\subsection{Tidal stream candidates for follow-up studies}
\label{sec:res_candidates}

\begin{figure*}
\centering
\renewcommand*{\arraystretch}{0}
\begin{tabular}{*{2}{@{}c}@{}}

\begin{overpic}[width=0.4\linewidth]{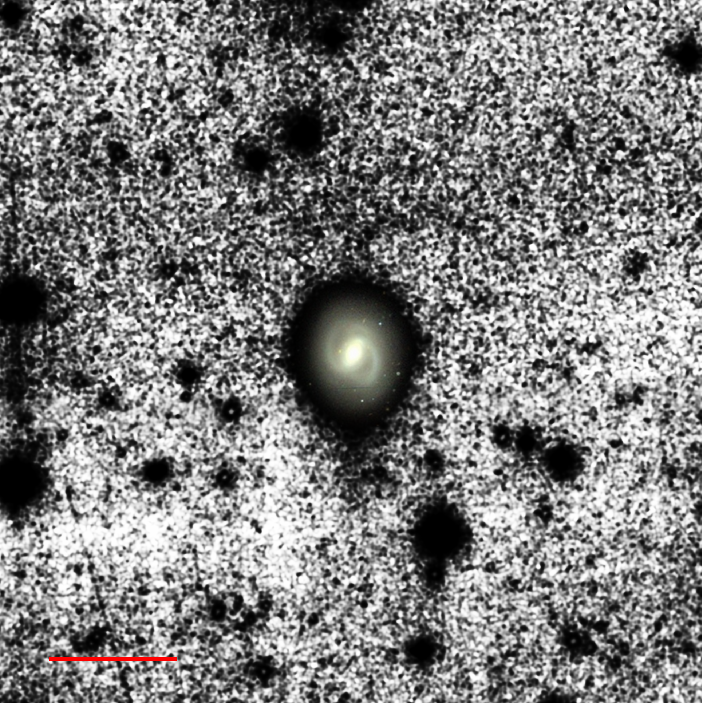}
\put(75,1){\colorbox{white}{\parbox{0.4\linewidth}{NGC 718}}}
\end{overpic}&
\begin{overpic}[width=0.4\linewidth]{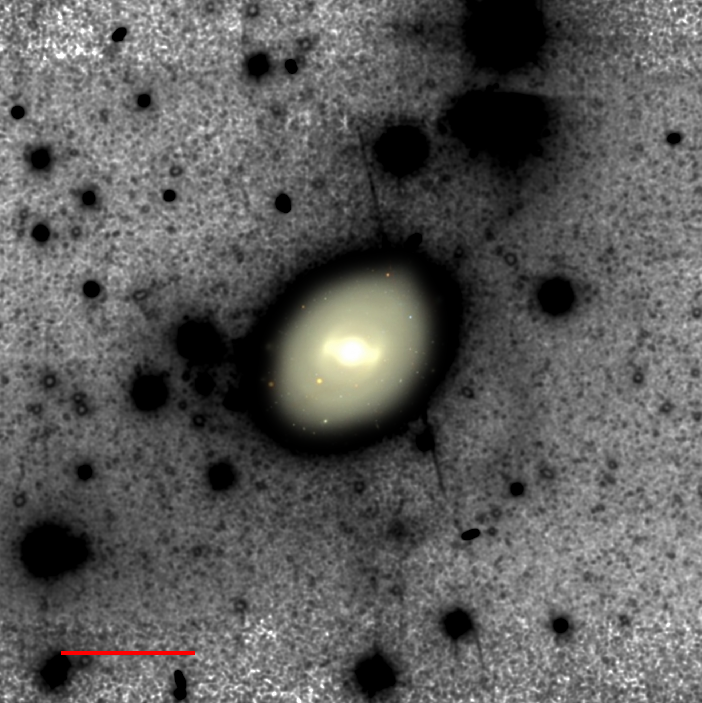}
\put(75,1){\colorbox{white}{\parbox{0.4\linewidth}{NGC 936}}}
\end{overpic}\\
\begin{overpic}[width=0.4\linewidth]{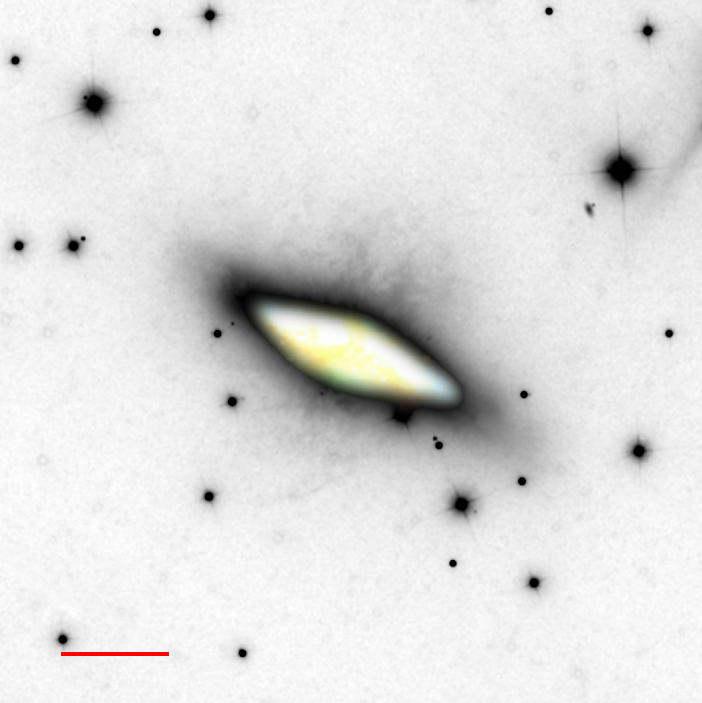}
\put(75,1){\colorbox{white}{\parbox{0.4\linewidth}{NGC 3034}}}
\end{overpic}&
\begin{overpic}[width=0.4\linewidth]{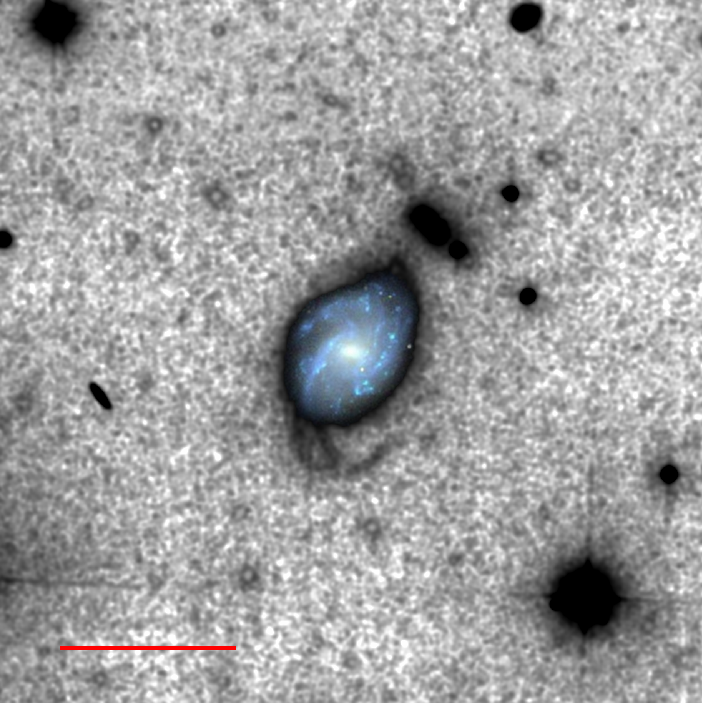}
\put(75,1){\colorbox{white}{\parbox{0.4\linewidth}{NGC 4519}}}
\end{overpic}\\
\begin{overpic}[width=0.4\linewidth]{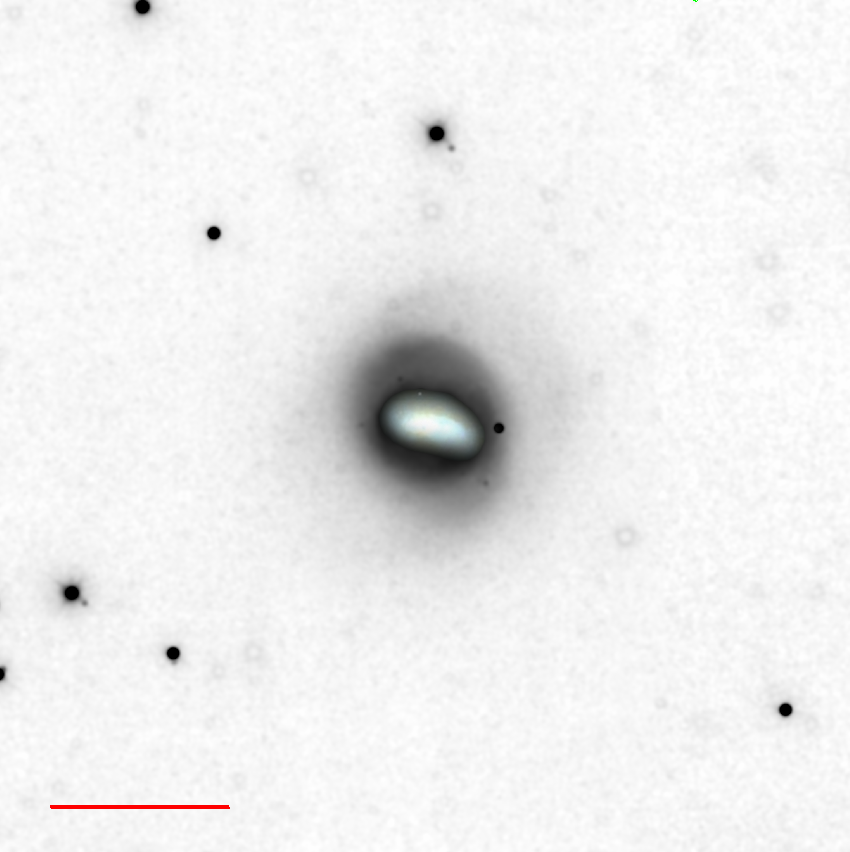}
\put(75,1){\colorbox{white}{\parbox{0.4\linewidth}{NGC 4691}}}
\end{overpic}&
\begin{overpic}[width=0.4\linewidth]{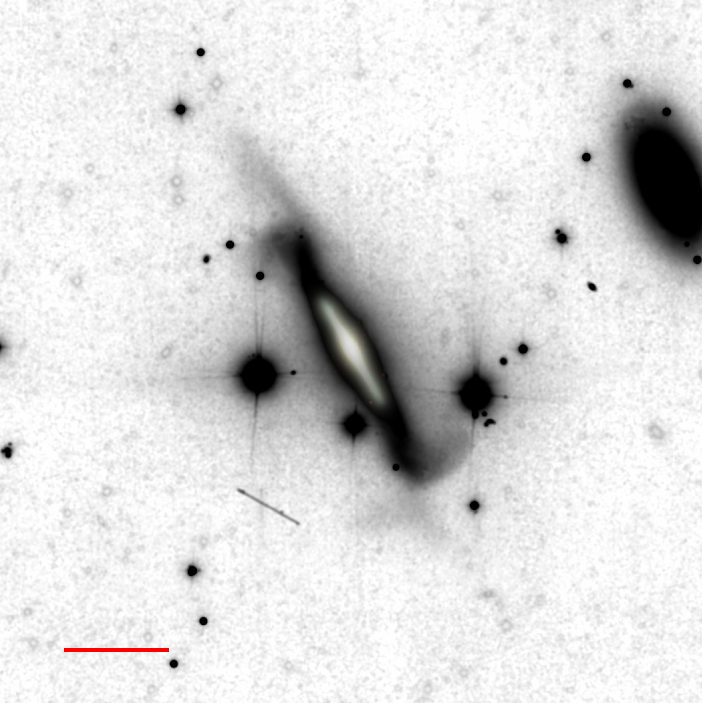}
\put(75,1){\colorbox{white}{\parbox{0.4\linewidth}{NGC 4762}}}
\end{overpic}\\
\end{tabular}
\caption{Selection of the diffuse-light features detected around some of the galaxies listed in Table \ref{tab:main_table}. Deeper data are needed to confirm their origin (tidal streams, galactic perturbations, extended spiral arms, etc). The red lines indicate a scale of 3 arcminutes. North is up and east to the left.}
\label{fig:candidates}
\end{figure*}

\begin{figure}
\centering
\renewcommand*{\arraystretch}{0}
\begin{tabular}{*{2}{@{}c}@{}}

\begin{overpic}[width=0.5\linewidth]{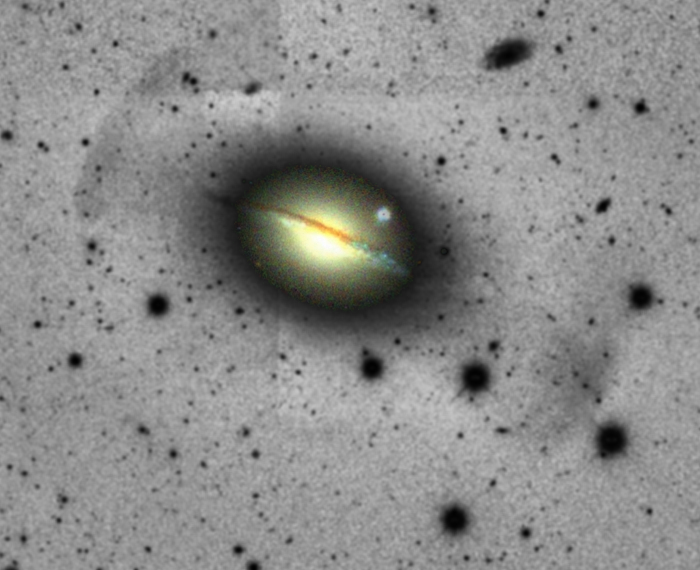}
\put(0,2){\colorbox{white}{\parbox{0.15\linewidth}{NGC 681}}}
\end{overpic}&
\begin{overpic}[width=0.5\linewidth]{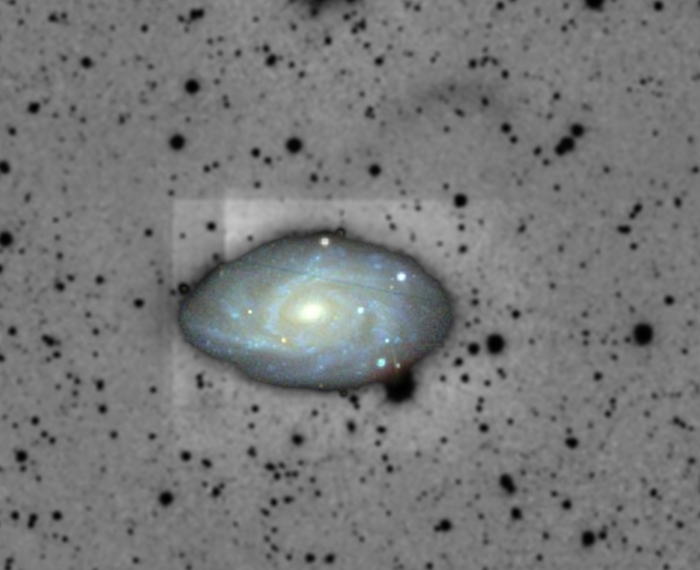}
\put(0,2){\colorbox{white}{\parbox{0.17\linewidth}{NGC 3041}}}
\end{overpic}\\
\begin{overpic}[width=0.5\linewidth]{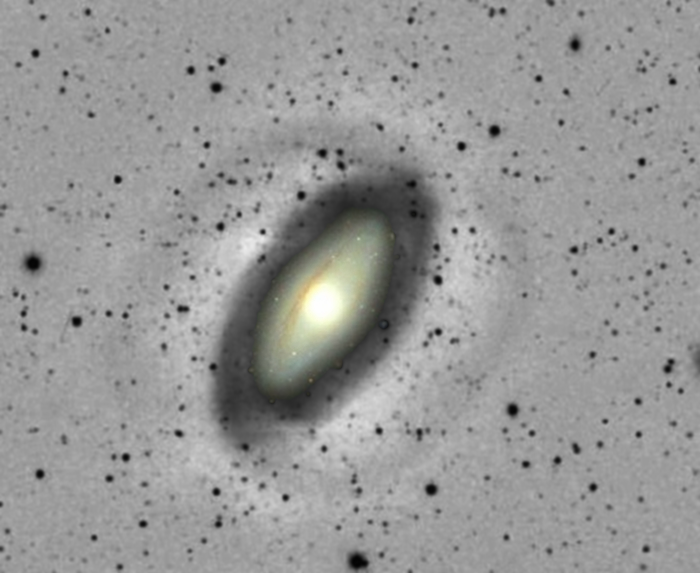}
\put(0,2){\colorbox{white}{\parbox{0.17\linewidth}{NGC 4772}}}
\end{overpic}&
\begin{overpic}[width=0.5\linewidth]{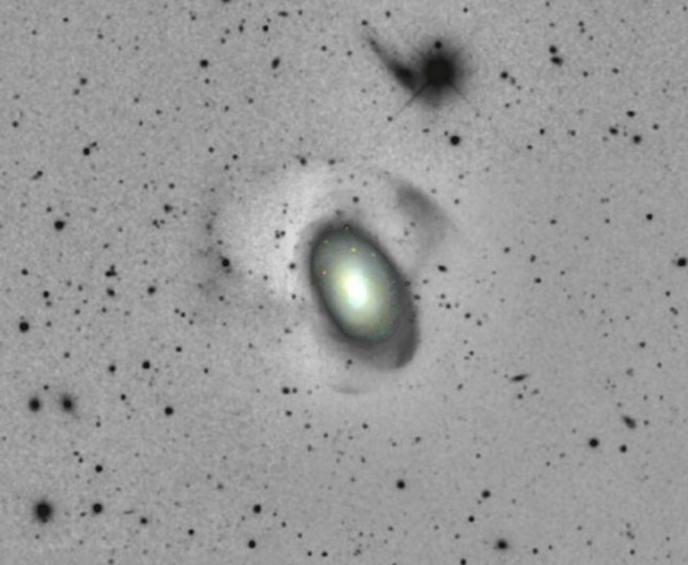}
\put(0,2){\colorbox{white}{\parbox{0.17\linewidth}{NGC 3611}}}
\end{overpic}\\
\begin{overpic}[width=0.5\linewidth]{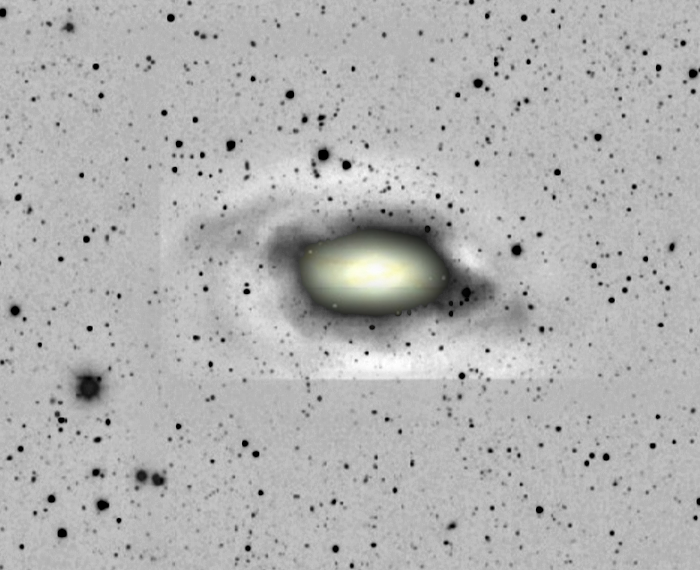}
\put(0,2){\colorbox{white}{\parbox{0.17\linewidth}{NGC 4753}}}
\end{overpic}&
\begin{overpic}[width=0.5\linewidth]{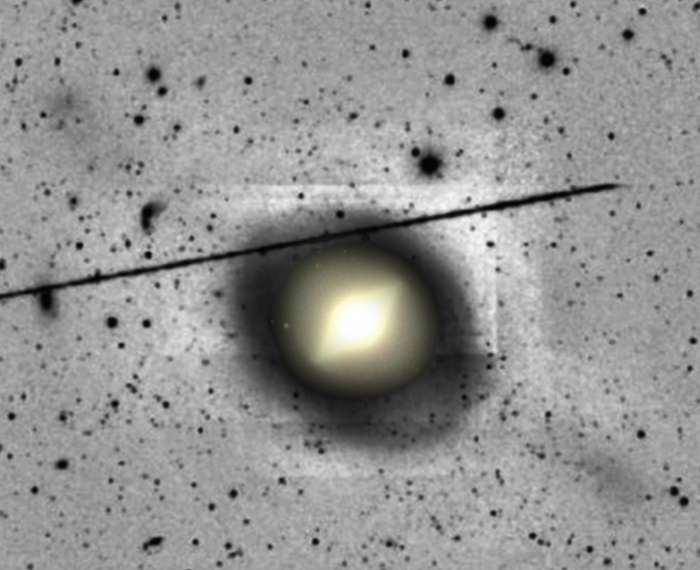}
\put(0,2){\colorbox{white}{\parbox{0.17\linewidth}{NGC 4643}}}
\end{overpic}\\
\end{tabular}
\caption{Images  taken from DECaLS public survey, confirming some of our findings. See Figure \ref{fig:confirmed} for more information. A color inset of the disk of each galaxy taken from this survey has been included as reference.}
\label{fig:decam_final}
\end{figure}

Table \ref{tab:main_table} lists the galaxies of our sample with detected structures for which a tidal origin cannot be confirmed in this work because we lack deeper data; $23$ in total, with a DCL equal to 1 or 2. Follow-up observations of these galaxies are currently being carried out by the STSS and will be published in a forthcoming companion contribution (Mart\'inez-Delgado et al., in preparation). These signatures define those that are very probably stellar tidal streams, that is, tracing orbits of satellites in the tidal field of the host galaxy, and features that are probably linked to disk warping, polar rings, and other types of signatures. In general, any features more likely related to galactic perturbations of the central galaxy disk due to dynamical interaction with other massive galaxies were tagged accordingly. Some examples of these structures of different types are displayed in Fig. \ref{fig:candidates}. Figure \ref{fig:decam_final} shows the images used to confirm the faint tidal debris detected around six of the galaxies listed in Table \ref{tab:main_table}.

\section{Conclusions}
\label{sec:con}

We have estimated the frequency of stellar tidal streams in the halos of massive galaxies in the local Universe by processing SDSS images to reveal low surface brigthness features, using a technique similar to that of \citet{miskolczi11}. Our results are summarized in Table \ref{tab:main_table}. To facilitate statistical comparisons with cosmological simulations of galaxy formation, we have defined a volume-, mass-, and size-limited parent sample of galaxies with stellar masses similar to that of the Milky Way based on the \sg\ catalog. From the $2331$ galaxies listed by \sg , our sample selects $297$ targets from the SDSS footprint (excluding low Galactic latitudes, major mergers, and the Virgo cluster). We estimate that the typical surface brightness limit of the SDSS images for these galaxies (after stacking their $g$, $r,$ and $i$ band images) is $28.1\ \pm 0.3\ \mathrm{mag~arcsec^{-2}}$.

By visual inspection, we  detected a total of $28$ confirmed tidal streams, including new features discovered in this study and some previously known tidal streams. Therefore, our most conservative estimate is that $9\%$ of the galaxies in our sample show evidence of diffuse features that may be linked to minor merging events (either stellar or gaseous streams, or a mixture of both). This fraction of galaxies displaying tidal features does not include the possible new, but unconfirmed detections listed in Table \ref{tab:main_table}. When we also count the systems with high-confidence detections (i.e., with a DCL of 3 or 4), the frequency of tidal features in our sample rises to $14\%$. It is important to remark that some of these  diffuse-light features may not be the signature of dwarf satellite remnants, but instead Galactic cirrus, imaging artifacts, or distorted spiral arms. This underscores the importance of deeper observations to confirm the nature of these features (see Fig. \ref{fig:examples}).

These results are broadly consistent with comparable studies cited in Section \ref{sec:intro}, in particular those listed in Table 1 from \citet{Atkinson2013}. Although the surface brightness limits of the observations used in these earlier studies are more or less compatible, the wide variety of sample selections limits a more detailed comparison of the final results. Bright tidal features are expected to be relatively more likely in ETGs, while the only analogous study for disk galaxies \citep{miskolczi11} was also less statistically representative of the galaxy population because it was focused on testing the image processing method. Comparisons with simulations are still needed, and will be reported elsewhere.

Last, because our procedure for enhancing images to detect low
surface brightness features relies on stacking images in multiple filters and because those features have an intrinsically low S/N, we cannot measure their colours. To do so, deeper multi-band imaging is needed, and will be valuable to constrain stellar populations and masses of the merging systems. Discussion of the physical properties of the features we detected and their comparison to the newest simulations is beyond the scope of this article. These topics will be addressed in a forthcoming paper.

%\begin{figure*}
%\includegraphics[width=\linewidth]{classes2.png}
%\caption{Different examples of overdensities that can be found. From left to right: a tidal stream, a diffuse feature not clearly linked to minor merger events (and hard to classify without better data), a confirmed image artifact, and a Galactic cirrus. Only %the first and second cases are the focus of this work.}
%\label{fig:examples}
%\end{figure*}

\begin{acknowledgements}
     GM thanks Arpad Miskolczi for the image processing training and the SDSS images showed in Fig. \ref{fig:misk_vs_stss}, and the Astronomisches Institut der Ruhr-Universitat Bochum for their hospitality during his visit. GM also thanks Gabriel Torrealba for his help in mapping the SDSS DR10 footprint, and N\'estor Espinoza for his comments on the statistics of the sample. DMD thanks the team of astrophotographers from the STSS: Mark Hanson, Adam Block and Ken Crawford for their images shown in this work; Giuseppe Donatiello for his help with the presentation of some figures; and Seppo Laine for his help with the target selection and comments. APC acknowledges the support of STFC Grant ST/P000541/1. We also thank the anonymous referee for their constructive comments in the revision stage.

     The authors acknowledge the use of the Spitzer Survey of Stellar Structure in Galaxies and the Dark Energy Camera Legacy Survey as a core part of this work.

     The authors also acknowledge the use of IRAF, the Image Reduction and Analysis Facility, a general purpose software system for the reduction and analysis of astronomical data. IRAF is written and supported by the National Optical Astronomy Observatories (NOAO) in Tucson, Arizona. NOAO is operated by the Association of Universities for Research in Astronomy (AURA), Inc. under cooperative agreement with the National Science Foundation.

     Funding for SDSS-III has been provided by the Alfred P. Sloan Foundation, the Participating Institutions, the National Science Foundation, and the U.S. Department of Energy Office of Science. %The SDSS-III web site is http://www.sdss3.org/. SDSS-III is managed by the Astrophysical Research Consortium for the Participating Institutions of the SDSS-III Collaboration including the University of Arizona, the Brazilian Participation Group, Brookhaven National Laboratory, Carnegie Mellon University, University of Florida, the French Participation Group, the German Participation Group, Harvard University, the Instituto de Astrofisica de Canarias, the Michigan State/Notre Dame/JINA Participation Group, Johns Hopkins University, Lawrence Berkeley National Laboratory, Max Planck Institute for Astrophysics, Max Planck Institute for Extraterrestrial Physics, New Mexico State University, New York University, Ohio State University, Pennsylvania State University, University of Portsmouth, Princeton University, the Spanish Participation Group, University of Tokyo, University of Utah, Vanderbilt University, University of Virginia, University of Washington, and Yale University.

     This work made use of the following Python 3 libraries: Matplotlib \citep{Hunter:2007}, SciPy \citep{scipy2001}, NumPy. This research also made use of Astropy, a community-developed core Python package for Astronomy \citep{Greenfield:2003}.

     %Part of this work was supported by %the German
      %\emph{Deut\-sche For\-schungs\-ge\-mein\-schaft, DFG\/} project
      %number Ts~17/2--1.
\end{acknowledgements}

% WARNING
%-------------------------------------------------------------------
% Please note that we have included the references to the file aa.dem in
% order to compile it, but we ask you to:
%
% - use BibTeX with the regular commands:
%   \bibliographystyle{aa} % style aa.bst
%   \bibliography{Yourfile} % your references Yourfile.bib
%
% - join the .bib files when you upload your source files
%-------------------------------------------------------------------
%\newpage
\bibliographystyle{aa} % style aa.bst
\bibliography{main.bib} % your references Yourfile.bib

\onecolumn
\begin{appendix}

\section{Supplementary tables.}
\label{app}

%%\input{tbl/table1_new.tex}
% Please add the following required packages to your document preamble:
% \usepackage{booktabs}
% \usepackage{longtable}
% Note: It may be necessary to compile the document several times to get a multi-page table to line up properly
%\newpage
\definecolor{Gray}{gray}{0.95}
\begin{longtable}[c]{@{}llccccl@{}}
\toprule
ID & \begin{tabular}[c]{@{}l@{}}Galaxy\\ morphology\end{tabular} & \multicolumn{1}{c}{\begin{tabular}[c]{@{}c@{}}Distance\\ (Mpc)\end{tabular}} & \multicolumn{1}{c}{\begin{tabular}[c]{@{}c@{}}log(Stellar Mass)\\ $\mathrm{(M_\odot)}$\end{tabular}} & DCL & Tags & Comments \\* \midrule
\endfirsthead
\multicolumn{7}{c}%
{Table \thetable\ continued from previous page} \\
\toprule
ID & \begin{tabular}[c]{@{}l@{}}Galaxy\\ morphology\end{tabular} & \multicolumn{1}{c}{\begin{tabular}[c]{@{}c@{}}Distance\\ (Mpc)\end{tabular}} & \multicolumn{1}{c}{\begin{tabular}[c]{@{}c@{}}log(Stellar Mass)\\ $\mathrm{(M_\odot)}$\end{tabular}} & DCL & Tags & Comments \\* \midrule
\endhead

NGC  681 & SAB    & 33.600 & 10.752 & 4 &     S C & This work\\
\rowcolor{Gray}
NGC  718 & SAB    & 21.400 & 10.283 & 3 &     C E & \begin{tabular}[c]{@{}l@{}}Very faint arc-like feature to the north\\ plus possible overdensity to the south\end{tabular}\\
NGC  936 & SB0    & 20.683 & 10.926 & 3 &     C E & \begin{tabular}[c]{@{}l@{}}Double arc-like feature;\\ hints of warped disk\end{tabular} \\
\rowcolor{Gray}
NGC 1055 & SBb    & 16.630 & 10.739 & 4 &       O & \citet{dmd10}\\
NGC 1084 & SAc    & 21.225 & 10.619 & 4 &       C & \citet{dmd10}\\
\rowcolor{Gray}
NGC 2775 & SAab   & 17.000 & 10.870 & 4 &       S & This work; MD+\\
NGC 2859 & (R)SB0 & 27.333 & 10.882 & 3 &   C Sph & \begin{tabular}[c]{@{}l@{}}Two possible partially disrupted satellites\\
within a ring, with leading and trailing tails\end{tabular}\\
\rowcolor{Gray}
NGC 3034 & I0     &  3.777 & 10.449 & 2 &     E O & Possible spike features\\
NGC 3041 & SAB    & 26.350 & 10.437 & 4 &       C & This work\\
\rowcolor{Gray}
NGC 3049 & SBab   & 30.775 & 10.132 & 4 &   S C E & This work; MD+\\
NGC 3185 & (R)SBa & 24.725 & 10.215 & 3 &   C E O & \begin{tabular}[c]{@{}l@{}}Very faint loop connected to the disk,\\ with a compact object embedded on it\end{tabular}\\
\rowcolor{Gray}
NGC 3277 & SA     & 25.000 & 10.375 & 1 &       S & Possible shells very close to the halo\\
NGC 3521 & SABbc  & 12.078 & 11.030 & 4 & S C E O & \citet{dmd10}\\
\rowcolor{Gray}
NGC 3611 & SAa    & 33.300 & 10.462 & 4 &       C & \begin{tabular}[c]{@{}l@{}}\citet{Schweizer1990};\\
This work\end{tabular}\\
NGC 3628 & Sb     & 11.300 & 10.805 & 4 &     Sph & \citet{dmd10}\\
\rowcolor{Gray}
NGC 3631 & SAc    & 13.102 & 10.163 & 4 &     E O & This work; MD+\\
NGC 3675 & SA     & 17.200 & 10.919 & 1 &       E & \begin{tabular}[c]{@{}l@{}}Candidate tidal overdensities, not\\ clearly distinguishable from disk warping\end{tabular}\\
\rowcolor{Gray}
NGC 3682 & SA0    &     ND & 10.230 & 4 &       S & This work\\
NGC 3729 & SB     & 20.183 & 10.233 & 3 &   Sph E & Possible satellite being disrupted\\
\rowcolor{Gray}
NGC 3877 & SA     & 15.612 & 10.445 & 1 &     E O & \begin{tabular}[c]{@{}l@{}}Asymmetrical and coplanar spike\\ extending from the disk\end{tabular}\\
NGC 3949 & SA     & 18.341 & 10.246 & 3 &     S E & Possible shell very close to the outer disk\\
\rowcolor{Gray}
NGC 4013 & Sb     & 18.600 & 10.630 & 4 &     E O & \citet{dmd10}\\
NGC 4051 & SAB    & 14.575 & 10.359 & 3 & Sph E O & \begin{tabular}[c]{@{}l@{}}Possible compact object with halo and tail,\\ plus an overdensity south of the galaxy\end{tabular}\\
\rowcolor{Gray}
NGC 4111 & SA0    & 15.550 & 10.452 & 4 &   Sph O & \citet{Brodie2014}\\
NGC 4203 & SAB0   & 14.940 & 10.528 & 4 &     Sph & This work\\
\rowcolor{Gray}
NGC 4262 & SB0    & 20.510 & 10.377 & 2 &       E & \begin{tabular}[c]{@{}l@{}}Two overdensities not clearly related to\\ tidal features, perhaps part of the disk\end{tabular}\\
NGC 4293 & (R)SB0 & 14.320 & 10.418 & 3 &     E O & \begin{tabular}[c]{@{}l@{}}Clear substructure in the inner halo,\\ very close to the disk\end{tabular}\\
\rowcolor{Gray}
NGC 4394 & (R)SB  & 16.800 & 10.440 & 3 &       E & \begin{tabular}[c]{@{}l@{}}Possible extended disk features,\\ or tidal arcs surrounding the galaxy\end{tabular}\\
NGC 4414 & SAc    & 18.312 & 10.883 & 4 &       S & \citet{deBlok2014}\\
\rowcolor{Gray}
NGC 4494 & E      & 13.841 & 10.542 & 2 &       O & \begin{tabular}[c]{@{}l@{}}Possible diffuse substructure, resembling\\ symmetric spikes in an elliptical galaxy\end{tabular}\\
NGC 4519 & SB     & 28.411 & 10.191 & 3 &     C E & \begin{tabular}[c]{@{}l@{}}Filamentary feature with two components,\\ likely related to either the halo or the disk\end{tabular}\\
\rowcolor{Gray}
NGC 4569 & SABab  & 12.352 & 10.638 & 4 &     E O & \citet{dmd10}\\
NGC 4594 & SAa    & 10.390 & 11.253 & 4 &       C & \citet{Malin97}\\
\rowcolor{Gray}
NGC 4631 & SBd    &  6.050 & 10.127 & 4 &     E O & \citet{dmd15}\\
NGC 4643 & SB0    & 25.700 & 11.028 & 4 &   C Sph & This work\\
\rowcolor{Gray}
NGC 4651 & SAc    & 26.708 & 10.844 & 4 & S C E O & \citet{dmd10}\\
NGC 4691 & (R)SB0 & 22.500 & 10.479 & 2 &     E O & \begin{tabular}[c]{@{}l@{}}Possible outer halo overdensity with\\ the appearance of a dense stellar cloud\end{tabular}\\
\rowcolor{Gray}
NGC 4753 & I0     & 16.869 & 10.930 & 4 &       E & \citet{steiman-cameron1992}\\
NGC 4762 & SB0    & 22.460 & 10.848 & 2 &     E O & \begin{tabular}[c]{@{}l@{}}An interesting case of disk warping\\ with mixed tidal features\end{tabular}\\
\rowcolor{Gray}
NGC 4772 & SAa    & 30.475 & 10.747 & 4 &       E & \citet{Haynes2000}\\
NGC 4866 & SA0    & 23.800 & 10.689 & 1 &     E O & \begin{tabular}[c]{@{}l@{}}Unclassifiable disk feature to the right\\ of the galaxy, possibly with tidal origin\end{tabular}\\
\rowcolor{Gray}
NGC 5055 & SAbc   &  8.333 & 10.778 & 4 &     C O & \citet{dmd10}\\
NGC 5364 & SA     & 19.513 & 10.614 & 3 &     E O & Giant tidal structure west of the galaxy\\
\rowcolor{Gray}
NGC 5506 & S pec  & 23.833 & 10.122 & 3 &     C O & \begin{tabular}[c]{@{}l@{}}Distorted, asymmetric tidal features\\ connected to each side of the disk\end{tabular}\\
NGC 5576 & E      & 23.930 & 10.770 & 1 &       S & Possible diffuse shells\\
\rowcolor{Gray}
NGC 5750 & SB0    & 33.633 & 10.741 & 4 &     E O & This work\\
NGC 5806 & SAB    & 25.541 & 10.585 & 3 &   C S E & \begin{tabular}[c]{@{}l@{}}Diffuse extended overdensity, with a shell\\ or arc-like feature very close to the disk\end{tabular}\\
\rowcolor{Gray}
NGC 5907 & SAc    & 16.636 & 10.871 & 4 &       C & \citet{dmd10}\\
NGC 7241 & SB     &     ND & 10.263 & 4 &     E O & \citet{Leaman2015}\\
\rowcolor{Gray}
NGC 7742 & SAb    & 22.200 & 10.343 & 4 &     S C & This work\\
NGC 7743 & (R)SB0 & 21.433 & 10.447 & 4 &     C E & This work; MD+\\*

\bottomrule
\caption{Tidal streams found in this work, including previously known features and new discoveries (28 host galaxies), with detection confidence levels (DCL) 3 and 4. Diffuse-light overdensities, with their physical nature yet to be confirmed (23 host galaxies) are also included, with DCL 1 and 2. This implies a total of 51 galaxies with any type of tidal features related to them. Distances and stellar masses were taken from \sg, while their morphology was extracted from NED database. Additionally, substructures we found have been tagged: \textit{S} for shells; \textit{C} for curved, arcuated features, including anything coherent and stream like; \textit{Sph} for spheroidal satellites and partially disrupted cores; \textit{E} for extensions of the central galaxy (e.g. warps and spiral arms); and \textit{O} for any other type of less common features (wedges, radial spikes, fuzzy clouds of debris, etc.). For known substructures, references of previous studies have been supplied. MD+ refers to Mart\'inez-Delgado et al. (in prep.), a forthcoming paper.}
\label{tab:main_table}\\
\end{longtable}

%%\input{tbl/table3.tex}
% Please add the following required packages to your document preamble:
% \usepackage{graphIC x}
\begin{table*}[h]
\centering
\resizebox{\textwidth}{!}{%
\begin{tabular}{|l|l|l|l|l|l|l|}
NGC 7814 & NGC 2967  & NGC 3495 & NGC 3992  & NGC 4343  & NGC 4904  & NGC 5792 \\
NGC 157  & NGC 2964  & NGC 3501 & IC 749    & NGC 4356  & NGC 5005  & NGC 5821 \\
NGC 337  & UGC 5228  & NGC 3507 & IC 750    & NGC 4369  & NGC 5033  & NGC 5854 \\
NGC 584  & NGC 3003  & NGC 3512 & NGC 4030  & NGC 4380  & NGC 5112  & NGC 5864 \\
NGC 615  & NGC 3021  & NGC 3556 & NGC 4045  & IC 3322A  & NGC 5145  & NGC 5879 \\
NGC 628  & NGC 3044  & NGC 3596 & NGC 4062  & UGC 7522  & NGC 5205  & NGC 5921 \\
NGC 660  & NGC 3055  & NGC 3623 & NGC 4085  & NGC 4405  & IC 902    & NGC 5963 \\
NGC 676  & NGC 3031  & NGC 3626 & NGC 4088  & NGC 4448  & UGC 8614  & NGC 5956 \\
NGC 693  & NGC 3067  & NGC 3629 & NGC 4096  & NGC 4451  & NGC 5248  & NGC 5957 \\
NGC 701  & NGC 3098  & NGC 3637 & NGC 4100  & NGC 4461  & NGC 5301  & NGC 5962 \\
NGC 779  & NGC 3162  & NGC 3642 & NGC 4102  & NGC 4503  & NGC 5300  & NGC 5964 \\
IC 210   & NGC 3177  & NGC 3655 & NGC 4123  & NGC 4437  & NGC 5334  & NGC 5970 \\
NGC 864  & NGC 3185  & NGC 3666 & NGC 4138  & NGC 4527  & NGC 5356  & NGC 6015 \\
NGC 955  & NGC 3184  & NGC 3669 & NGC 4145  & NGC 4536  & NGC 5422  & NGC 6012 \\
NGC 1022 & NGC 3198  & NGC 3681 & NGC 4157  & NGC 4559  & NGC 5443  & IC 1158  \\
NGC 1035 & IC 610    & NGC 3684 & UGC 7267  & NGC 4565  & NGC 5457  & NGC 6106 \\
UGC 4551 & NGC 3254  & NGC 3683 & NGC 4212  & NGC 4580  & NGC 5473  & NGC 6217 \\
NGC 2654 & NGC 3259  & NGC 3686 & NGC 4217  & NGC 4599  & NGC 5480  & NGC 7280 \\
NGC 2683 & NGC 3279  & NGC 3692 & NGC 4220  & NGC 4632  & NGC 5520  & NGC 7497 \\
NGC 2712 & NGC 3294  & NGC 3755 & NGC 4237  & NGC 4639  & NGC 5507  & NGC 7625 \\
NGC 2742 & NGC 3346  & NGC 3756 & NGC 4260  & NGC 4666  & NGC 5584  & NGC 1052 \\
NGC 2770 & NGC 3351  & IC 719   & NGC 4274  & NGC 4710  & NGC 5668  & NGC 2768 \\
NGC 2780 & NGC 3359  & NGC 3810 & UGC 7387  & NGC 4746  & NGC 5690  & NGC 3193 \\
NGC 2805 & NGC 3370  & NGC 3900 & NGC 4303  & NGC 4771  & NGC 5713  & NGC 3608 \\
NGC 2820 & NGC 3389  & NGC 3898 & NGC 4307  & NGC 4800  & IC 1048   & NGC 4278 \\
NGC 2844 & NGC 3430  & NGC 3938 & NGC 4314  & NGC 4808  & NGC 5746  & NGC 5173 \\
NGC 2841 & NGC 3437  & NGC 3953 & NGC 4316  & NGC 4826  & NGC 5768  & NGC 5216 \\
NGC 2903 & NGC 3486  & NGC 3982 & NGC 4324  & NGC 4845  & NGC 5798  & NGC 5846
\end{tabular}%
}
\caption{Galaxies with no evidence of observable diffuse overdensities in our sample of 297 galaxies.}
\label{tab:negatives}
\end{table*}

\end{appendix}

\end{document}